\documentclass[10pt,a4paper,american,superscriptaddress,aps,prd,nofootinbib]{revtex4}
\usepackage{lmodern}

\usepackage[T1]{fontenc}
\usepackage[utf8]{inputenc}
\usepackage{amsmath}
\usepackage{amsfonts}
\usepackage{amssymb}
\usepackage{graphicx}
\usepackage{bm}
\usepackage{esint}
\usepackage{multirow}
\usepackage{color}
\usepackage{babel}

\makeatletter

\providecommand{\tabularnewline}{\\}

\usepackage{hyperref}\hypersetup{
    colorlinks,%
    citecolor=blue,%
    filecolor=blue,%
    linkcolor=blue,%
    urlcolor=blue
}



\begin{document}

\title{Model-independent constraints on the cosmological anisotropic stress}

\author{Luca Amendola}

\affiliation{Institut Für Theoretische Physik, Ruprecht-Karls-Universität Heidelberg,
Philosophenweg 16, 69120 Heidelberg, Germany}

\author{Simone Fogli}

\affiliation{Institut Für Theoretische Physik, Ruprecht-Karls-Universität Heidelberg,
Philosophenweg 16, 69120 Heidelberg, Germany}

\affiliation{Dipartimento di Fisica e Astronomia, Universit\`a di Bologna, Via Irnerio 46, 40126 Bologna, Italy}

\author{Alejandro Guarnizo}

\affiliation{Institut Für Theoretische Physik, Ruprecht-Karls-Universität Heidelberg,
Philosophenweg 16, 69120 Heidelberg, Germany}

\author{Martin Kunz}

\affiliation{Département de Physique Théorique and Center for Astroparticle Physics,
Université de Genéve, Quai E.\ Ansermet 24, CH-1211Genéve 4, Switzerland}

\affiliation{African Institute for Mathematical Sciences, 6 Melrose Road, Muizenberg,
7945, South Africa}

\author{Adrian Vollmer}

\affiliation{Institut Für Theoretische Physik, Ruprecht-Karls-Universität Heidelberg,
Philosophenweg 16, 69120 Heidelberg, Germany}

\begin{abstract}
The effective anisotropic stress or gravitational slip $\eta=-\Phi/\Psi$ is
a key variable in the characterisation of the physical origin of the dark
energy, as it allows to test for a non-minimal coupling of the dark sector
to gravity in the Jordan frame. It is however important to use a fully
model-independent approach when measuring $\eta$ to avoid introducing a
theoretical bias into the results. In this paper we forecast the precision
with which future large surveys can determine $\eta$ in a  way that only
relies on directly observable quantities. In particular, we do not assume
anything concerning the initial spectrum of perturbations, nor on its
evolution outside the observed redshift range, nor on the galaxy bias. We
first leave $\eta$ free to vary in space and time and then we model it as
suggested in Horndeski models of dark energy. Among our results, we find
that a future large scale lensing and clustering survey can constrain $\eta$
to within 10\% if $k$-independent, and to within 60\% or better at $k=0.1
h/$Mpc if it is restricted to follow the Horndeski model.
\end{abstract}

\date{\today}

\maketitle

\section{Introduction}\label{sec:int}

With the recent first results of the Planck satellite \cite{Ade:2013ktc}
we have definitely reached the era of precision cosmology: The Planck
observations of the cosmic microwave background (CMB) are well described
by the six-parameter flat $\Lambda$CDM model, and most of those six
parameters are determined to percent-level accuracy \cite{Ade:2013zuv}.
The most impressive achievement is the measurement of the acoustic
scale of the CMB with a precision of 0.06\% by Planck, but also the
physical baryon and the matter densities have been determined to within
an uncertainty of only 1 to 2\%.

But the conclusion from these measurements is that we live in an Universe
where only 5\% of today's energy density consists of the kind of matter
described by the standard model of particle physics. Another 27\%
appears to be matter that is only interacting gravitationally with
the visible world, and the remaining 68\% is made up of a cosmological
constant.

The physical nature of the dark sector is however completely unknown,
and especially the cosmological constant suffers from severe theoretical
problems. For this reason it is of crucial importance to look beyond
the perfectly homogeneous cosmological constant and to investigate
general dark energy models, including also modifications of Einstein's
theory of General Relativity (GR). When considering a general dark
energy model however, high precision is much harder to achieve, and
it is important to understand first what can actually be observed, to
avoid introducing a theoretical bias into the observational results.
Coming from this angle, we determined in a recent paper \cite{Amendola:2012ky}
that cosmological measurements at linear scales can determine, in addition to the expansion
rate $H(z)$, only three additional variables $R$, $A$ and $L$,
given by 
\begin{align}
A & =Gb\delta_{\text{m,0}}\,,\qquad R=Gf\delta_{\text{m,0}}\,,\label{eq:DirectObs}\\
L & =\Omega_{\text{m,0}}GY(1+\eta)\delta_{\text{m,0}}\,.\nonumber 
\end{align}
 Denoting with $k$ the norm of the wavenumber and with $a$ the cosmic scale factor, we refer with   $G(k,a)$ 
 to the linear growth function (normalized to unity today)
 with $f=G'/G$ to the growth rate, with
$b(k,a)$ to the galaxy bias with respect to the dark matter density
contrast and with $\delta_{\text{m,0}}(k)$ to the dark matter density
contrast today. The functions $\eta(k,a)$ and $Y(k,a)$ describe
the impact of the dark energy on the cosmological perturbations. Later on, we will also need
the quantities $\bar{A}\equiv A/\delta_{\text{t,0}}$,
$\bar{R}\equiv R/\delta_{\text{t,0}}$, $\bar{L}\equiv L/\delta_{\text{t,0}}$ 
with $\delta_{\text{t,0}} = \delta_{\text{m,0}}/\sigma_8$.  If we write the line element describing the
perturbed Friedmann-Lema\^{i}tre-Robertson-Walker metric as
\begin{equation}
ds^{2}=-(1+2\Psi){\mathrm d t}^{2}+a(t)^{2}\left(1+2\Phi\right){\mathrm d {\mathbf{x}}}^{2},\label{eq:metric}
\end{equation}
 then $\eta$ and $Y$ are defined through \cite{Amendola:2007rr,DeFelice:2011hq}
\begin{equation}
\eta(k,a)\equiv-\frac{\Phi}{\Psi}\,,\qquad
Y(k,a)\equiv-\frac{2k^{2}\Psi}{3\Omega_{\text{m}}\delta_{\text{m}}}\,. \label{eq:eta}
\end{equation}
We see that $\eta$ corresponds to the gravitational slip, which
is linked to the effective anisotropic stress of the dark energy,
and $Y$ describes the clustering of the dark energy. The function $\eta$
is particularly important, as it is a key-variable to distinguish scalar-field
type dark energy models from modifications of GR \cite{Saltas:2010tt,Sawicki:2012re}.

So far these are rather abstract considerations. An obviously important
question is whether we can actually measure these quantities with
realistic surveys, and to what precision. In \cite{Amendola:2012ky,Motta:2013cwa}
we showed that we can use the motion of light and of non-relativistic
test-particles like galaxies to map out the metric functions $\Phi$
and $\Psi$ in principle, and that therefore $\eta$ is an observable
quantity. But $Y$ depends on the dark matter distribution, which
is not directly observable, and so also $Y$ itself is in general
not directly observable due to the dark degeneracy \cite{Kunz:2007rk}.

In order to reconstruct $\eta$ from $A$, $R$ and $L$ it is necessary
to remove the dependence  on $\delta_{\text{t},0}$
(notice that $\bar A$, $\bar R$ and $\bar L$ are not observables), since it is an unknown quantity that does
not depend on dark energy physics but rather on inflation or other primordial effects.
This can be done by considering ratios like $P_{1}=R/A$, $P_{2}=L/R$
and $P_{3}=R'/R$. In terms of these model-independent ratios, the
gravitational slip becomes \cite{Amendola:2012ky,Motta:2013cwa} 
\begin{equation}
1+\eta=\frac{3P_{2}(1+z)^{3}}{2E^{2}\left(P_{3}+2+\frac{E'}{E}\right)} \label{eq:obseta}
\end{equation}
where we also set $E(z)\equiv H(z)/H_{0}$.

When constraining $\eta$ later on, we will use an equivalent quantity which we
call $\bar \eta$, defined as
\begin{equation}
\bar \eta \equiv \frac{2}{1+\eta} = \frac{2 \Psi}{\Psi-\Phi} \, .
\end{equation}
The reason is that even for large future surveys the expected error on $P_3$ is
substantial, especially when we want to allow for an unknown redshift and scale dependence.
The large error makes the division by $(P_3+2+E'/E)$ in Eq.\ (\ref{eq:obseta}) badly behaved.
$\bar\eta$ on the other hand is more stable, as we discuss in more detail in appendix \ref{sec:appeta}.

Based on these results, we will use the Fisher matrix formalism in this
paper to forecast the expected precision on $\bar A$, $\bar R$ and $\bar L$,
which are then projected onto the accuracy with which we can obtain $P_{1}$,
$P_{2}$ and $P_{3}$, and finally on $\bar \eta$, based on the expected performance of future large-scale galaxy
and weak lensing surveys. We will also include a supernova survey to improve
the constraints on the background expansion rate $E(z)$, although we find that its impact on the
final constraints on $\eta$ is rather modest. In
the final step we will assume four models for $\eta$: 
\begin{enumerate}
\item First, we assume that $\eta$ is constant at all scales and at all
redshifts (let us call this case the constant-$\eta$ case). This occurs
for instance in $\Lambda$CDM and in all models in which dark energy
does not cluster and is decoupled from gravity. 
\item Second, we assume that $\eta$ is constant in space but varies in
redshift ($z$-varying case). In other words, we assume that $\eta$
has a different arbitrary value for each redshift bin. 
\item Third, we assume $\eta$ varies in both redshift and space ($z,k$-varying
case). 
\item Fourth, we take for $\eta$ the quasi-static Horndeski result \cite{Amendola:2012ky}
\begin{equation}
\eta=h_{2}\left(\frac{1+k^{2}h_{4}}{1+k^{2}h_{5}}\right).\label{eq:obsetak}
\end{equation}
(Here we assume $k$ to be measured in units of 0.1 $h/$Mpc, so the $h_i$ functions are dimensionless). 
We denote this model as the Horndeski case. The Horndeski Lagrangian
is the most general Lagrangian for a single scalar field leading to
second-order equations of motion. The expression (\ref{eq:obsetak})
arises in the quasi-static limit \cite{DeFelice:2011hq} where the
time-derivative terms are sub-dominant, which implies that the scales
of interest are inside the (sound-) horizon. 
\end{enumerate}
In all cases the fiducial model will be chosen to be $\Lambda$CDM,
for which $\eta=\bar\eta=1$. For the first two cases we need only a binning
in redshift, while for the third and fourth case we will bin both
in redshift and in $k$-space. The fiducial values in the first Horndeski case are
$h_2=1$, $h_4=h_5=0$.

The outline of the paper is as follows: In sections \ref{sec:gc},
\ref{sec:wl} and \ref{sec:sn} we set up the Fisher matrix formalism for the
galaxy clustering, weak lensing, and SN-Ia observations. As already
mentioned above, we will see that we need to combine the different probes to
obtain constraints on $\eta$,  and we discuss the combination of the
Fisher matrices in Sec.~\ref{sec:comb} before concluding in the final
section.

\section{Notation and general definitions\label{sec:gendef}}

In this section we complete the definition of our notation and provide
definitions for quantities that are useful in several of the following
sections. Our metric signature and the gravitational potentials are
already defined in Eq.\ (\ref{eq:metric}). In Eq.\ (\ref{eq:eta})
we define the functions $\eta$ and $Y$ that parameterize the `dark
energy perturbations' (as the dark matter does not contribute to the
anisotropic stress\footnote{Beyond first order in perturbation theory, the dark matter
does in principle contribute to the pressure and anisotropic stress in the Universe,
but the contribution is very small and negligible for our purpose \cite{Ballesteros:2011cm}.}). 
The function $\eta$ assumes a central stage in this paper as
it is observable without requiring further assumptions, see Eq.\ (\ref{eq:obseta}).

Although the observables $E$, $A$, $R$ and $L$ can be measured
in a fully model-independent way, the precision with which we can
determine them depends also on the true nature of the Universe. When evaluating
our forecasts, we will use a flat $\Lambda$CDM fiducial model, characterized
by the WMAP 7-year values, $\Omega_{\text{m},0}h^{2}=0.134$, $\Omega_{b,0}h^{2}=0.022$,
$n_{s}=0.96$, $\tau=0.085$, $h=0.694$ and $\Omega_{k}=0$. The
new WMAP 9-year and Planck results are not very different so the
results are not significantly affected by our choice. The dimensionless
background expansion rate in the fiducial model and at low redshifts is given by 
\begin{equation}
E(z)^{2}=\Omega_{\text{m},0}(1+z)^{3}+(1-\Omega_{\text{m},0})\,,
\end{equation}
 and we will often use the dimensionless angular diameter distance
$\hat{d}_{A}(z)=\hat{r}(z)/(1+z)$ and the dimensionless luminosity
distance $\hat{d}_{L}(z)=\hat{r}(z)(1+z)$, where in a flat FLRW Universe
\begin{equation}
\hat{r}(z)=\int_{0}^{z}\frac{{\mathrm d \tilde{z}}}{E(\tilde{z})}\,.\label{eq:dist}
\end{equation}
 The usual distances are related to the dimensionless distances through
$\hat{r}=H_{0}r$ and $\hat{d}=H_{0}d$. In $\Lambda$CDM we have that
$\eta = 1$ and $Y=1$. In the fiducial model, both $G$ and $f$ only depend on the scale factor,
not on $k$. 

We will combine in the following the Fisher matrices for future galaxy
clustering, weak lensing and supernovae surveys. More specifically,
we will take for galaxy clustering (GC) and weak lensing (WL) a stage
IV kind of survey \cite{Albrecht:2006um} like Euclid%
\footnote{\url{http://www.euclid-ec.org/}%
} \cite{2011arXiv1110.3193L}. Notice that the survey specifications we use in this paper are meant only to be representative of a 
future dark energy survey and do not necessarily reflect the actual Euclid configuration.
 For supernovae (SN) we assume a survey of
$10^5$ sources with magnitude errors similar to the currently achievable
uncertainties, as expected in the  LSST survey \cite{Tyson:2003kb}.


\section{Galaxy clustering\label{sec:gc}}

The galaxy power spectrum can be written as \cite{Seo:2003} 
\begin{equation}
P(k,\mu)=(A+R\mu^{2})^{2}e^{-k^{2}\mu^{2}\sigma_{r}^{2}}=(\bar{A}+\bar{R}\mu^{2})^{2}\delta_{\text{t,0}}^{2}(k)e^{-k^{2}\mu^{2}\sigma_{r}^{2}},\label{gcpower}
\end{equation}
 where  $\sigma_{r}=\delta z/H(z)$,
$\delta z$ being  the absolute error on redshift measurement, and
we explicitly use $\delta_{\text{m,0}}=\sigma_{8}\delta_{\text{t,0}}$, and where
 $\mu$  is the cosine of the angle between the line of sight and the wavevector.
Notice that $\bar{R}$ is often denoted in the literature as $f\sigma_{8}(z)$.

As already emphasized, we will ignore in the
following the information contained in $\delta_{\text{t,0}}^{2}(k)$ since
this depends on initial conditions that are in general not known, and we
cannot disentangle the initial conditions from the information on the dark
energy (we refer to \cite{Amendola:2012ky} for a discussion about this
point). Removing the information
from the shape of the power spectrum of course reduces the amount of
information available and so increases the error bars. This is the price to
pay if we want to stay fully model independent.

The dependence on $E$ is implicitly contained in $\mu$ and $k$
through the Alcock-Paczyński effect \cite{Alcock:1979}.
However, we can only take into account the $\mu$ dependence, since the $k$ dependence
occurs through the unknown function $\delta_{\text{m,0}}$.
The Fisher matrix for the parameter vector $p_{\alpha}$ is in general \cite{Seo:2003} 
\begin{equation}
F_{\alpha\beta}^{\text{GC}}=\frac{1}{8\pi^{2}}\int_{-1}^{1}{\mathrm d \mu}\int_{k_{\text{min}}}^{k_{\text{max}}}k^{2} V_{\text{eff}} D_{\alpha}D_{\beta}\,{\mathrm d k} \, ,
\label{eq:fmgc}
\end{equation}
 where 
\begin{equation}
D_{\alpha}\equiv\frac{d\log P}{dp_{\alpha}}\biggl|_{r}
\end{equation}
is the parameter derivative evaluated on the fiducial values (designated by the subscript `$r$') and where
\begin{equation}
V_{\text{eff}}=\left(\frac{\bar{n}P(k,\mu)}{\bar{n}P(k,\mu)+1}\right)^{2} V_{\text{survey}}\label{veff}
\end{equation}
is the effective volume of the survey, with $\bar{n}$ the galaxy
number density in each bin (discussed later). The Fisher matrix is
evaluated at the fiducial model. 
For this evaluation we will assume that the bias in $\Lambda$CDM is scale
independent and equal to unity, which implies that the barred variables $\bar A$ and $\bar R$
also do not depend on $k$ in the fiducial model (although of course in
general they will be scale dependent). 

Our parameters are therefore $p_{\alpha}^{\text{\tiny{GC}}} = \{\bar{A}(\bar
z_1),\bar{R}(\bar z_1),E(\bar z_1),\bar{A}(\bar
z_2),\bar{R}(\bar z_2),E(\bar z_2), \dots\}$, where the subscripts run over the $z$ bins.   We could have used $A,R$ directly as parameters as in
Eq.~(\ref{gcpower}), but we prefer to clearly distinguish between the dark
energy dependent parameters $\bar{A}, \bar{R}$ and those that depend on
different physics. Indices $\alpha$ or $\beta$ always label the parameters in the Fisher matrix framework.
From the definition of the galaxy clustering power spectrum, Eq.~(\ref{gcpower}),
(and without taking into account the correction from the error on redshift, as we
will assume a spectroscopic survey with negligible redshift
errors) we find that\footnote{The simplicity of the angular dependence of these expressions and the relative insensitivity of the effective volume, 
Eq.\ (\ref{veff}) to $\mu$, mean that the Fisher matrix (\ref{eq:fmgc}) leads to a
generic prediction for galaxy clustering surveys: The measurements of $\bar A$ and $\bar R$ will be slightly anti-correlated, 
and galaxy clustering surveys can always measure $\bar A$ about 3.5 to 4.5 times better than $\bar R$.}
\begin{equation}
D_{\bar{A}} =\frac{2}{\bar{A}+\bar{R}\mu^{2}} \, ,  \quad D_{\bar{R}} =\frac{2\mu^{2}}{\bar{A}+\bar{R}\mu^{2}} \, ,\label{eq:DAR}
\end{equation}
 and using \cite[p.~393]{Amendola2010} 
\begin{equation}
\mu=\frac{H\mu_{r}}{H_{r}Q},
\end{equation}
 where 
\begin{equation}
Q = \frac{\sqrt{E^{2}\hat{d}_{A}^{2}\mu_{r}^{2}-E_{r}^{2}\hat{d}_{Ar}^{2}(\mu_{r}^{2}-1)}}{E_{r}\hat{d}_{A}},
\end{equation}
 we get for the derivative with respect to the parameter $E$ 
\begin{equation}
D_{E}=\frac{4\bar{R}\mu^{2}(1-\mu^{2})}{(\bar{A}+\bar{R}\mu^{2})}\left(\frac{1}{E_{r}}+\frac{1}{\hat{d}_{Ar}}\frac{\partial \, \hat{d}_{A}}{\partial E}\right).\label{DerE}
\end{equation}
Here we explicitly consider the dependence of the dimensionless
angular diameter distance $\hat{d}_{A}$ on $E$ via Eq.~(\ref{eq:dist}).

\subsection{$z$ binning}

\label{subsec:gcz}

We consider an Euclid-like survey \cite{2011arXiv1110.3193L} from
$z=0.5-1.5$ divided in equally spaced bins of width $\Delta z=0.2$, and, in
order to prevent accidental degeneracies due to low statistics, a single
larger redshift bin between $z=1.5-2.1$ (thus the number of bins is
$n_{B}=6$). 
The lower boundaries of the $z$-bins are labeled as $z_a$ while
the center of the bins are labeled as $\bar z_a$ (latin indices $a,b,\ldots$
label the $z$-bins).  The galaxy number densities in each bin are shown in
Table \ref{tab:pamsGC1}; for the bin between $1.5$ and $2.1$ we use an
average number of $0.33\times10^{-3}$ ($h/$Mpc)$^{3}$
\cite{Amendola:2012ys}. The error on the measured redshift is assumed to be
spectroscopic: $\delta z=0.001(1+z)$. The transfer function in the present
matter power spectrum ($\delta_{\text{t,0}}^{2}$) is calculated using CAMB
\cite{Lewis:2000} for the $\Lambda$CDM cosmology defined in
Sec.~\ref{sec:gendef}. The limits on the integration over $k$ are taken as
$k_{\text{min}}=0.007$ $h/$Mpc (but the results are very weakly dependent on
this value) and the values of $k_{\text{max}}$ are chosen to be well below
the scale of non-linearity at the redshift of the bin\footnote{The values of
$k_{\text{max}}$ are calculated imposing $\sigma^2(R)=0.35$, at the
corresponding $R=\pi/2k$ for each redshift, being $R$ the radius of spherical cells, see  \cite{Seo:2003}.}, see
Table~\ref{fid3}.

\begingroup 
\squeezetable
\begin{table}[b]
\begin{tabular}{ccccc}
\hline \hline
$\bar{z}$  & $k_{\text{min}}$  & $k_{1}$  & $k_{2}$  & $k_{\text{max}}$ \tabularnewline
\hline 
0.6  & \; 0.007 \;  & \; 0.022 \;  & \; 0.063 \;  & \; 0.180 \; \tabularnewline
0.8  & 0.007  & 0.023  & 0.071  & 0.215 \tabularnewline
1.0  & 0.007  & 0.024  & 0.078  & 0.249 \tabularnewline
1.2  & 0.007  & 0.026  & 0.086  & 0.287 \tabularnewline
1.4  & 0.007  & 0.027  & 0.094  & 0.329 \tabularnewline
1.8  & 0.007  & 0.029  & 0.112  & 0.426 \tabularnewline
\hline \hline
\end{tabular}\caption{\label{fid3}Values of $k_{1}$, $k_{2}$ and $k_{\text{max}}$ for
every redshift bin, in units of  ($h/$Mpc).}
\end{table}
\endgroup

Since the angular diameter distance can be approximated by the expression
\begin{equation}
\hat{d}_{A}(\bar{z}_a)=\frac{1}{(1+\bar z_{a})}\sum_{b=0}^{b=a}\frac{\Delta z_{b}}{E(\bar z_b)},
\end{equation}
we have for the term $\frac{\partial \,\hat{d}_{A}}{\partial E}$ in equation (\ref{DerE})
\begin{equation}
\frac{\partial\hat{d}_{A}(\bar{z}_a)}{\partial E(\bar z_{b})}=-\frac{\Delta z_{b}}{ (1+\bar{z}_{a})E_{b}^{2}}\delta_{ab},
\end{equation}
where $\delta_{ab}$ is a Kronecker delta symbol. Then we calculate the Fisher
matrix block-wise with independent submatrices $F_{\alpha\beta}^{\text{GC}}$
for each bin.

\begin{figure}[t]
\centering
\includegraphics[width=0.32\linewidth]{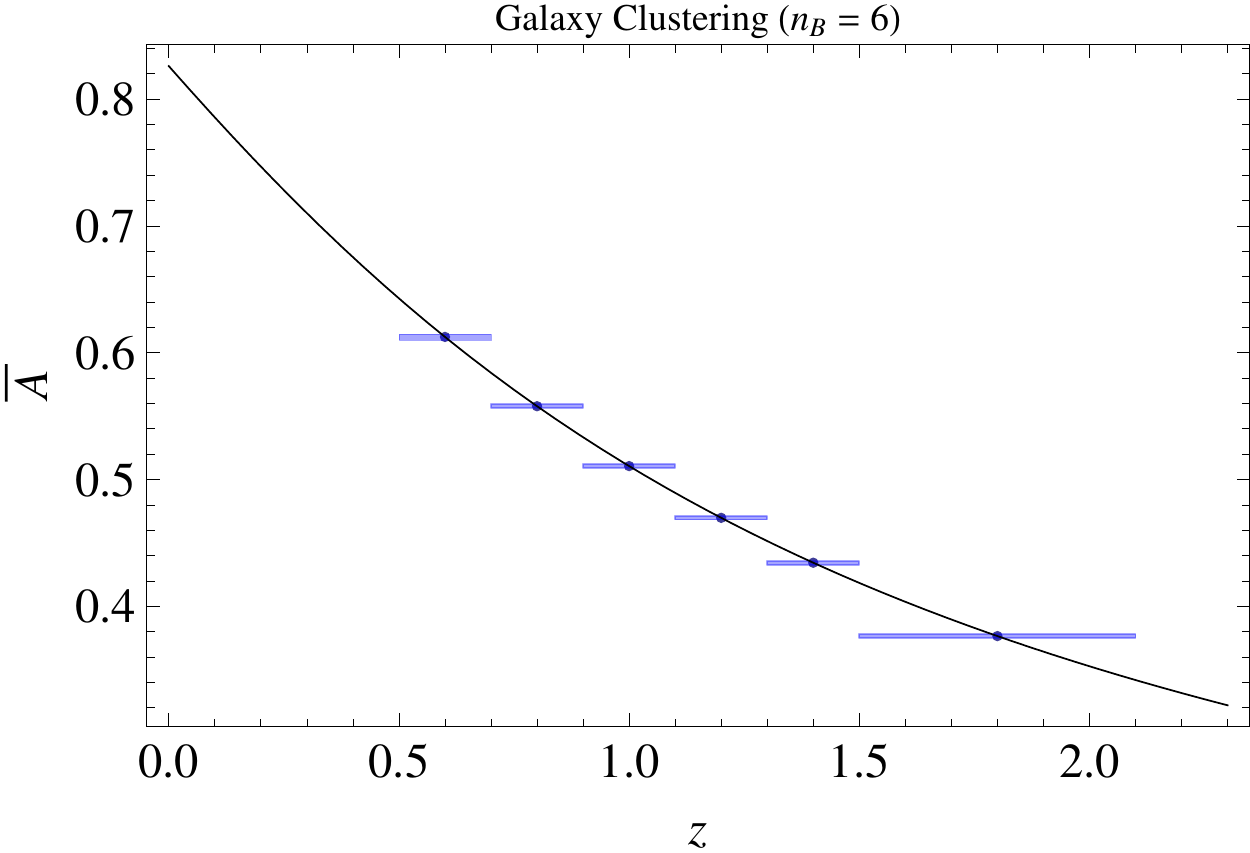} 
\includegraphics[width=0.32\linewidth]{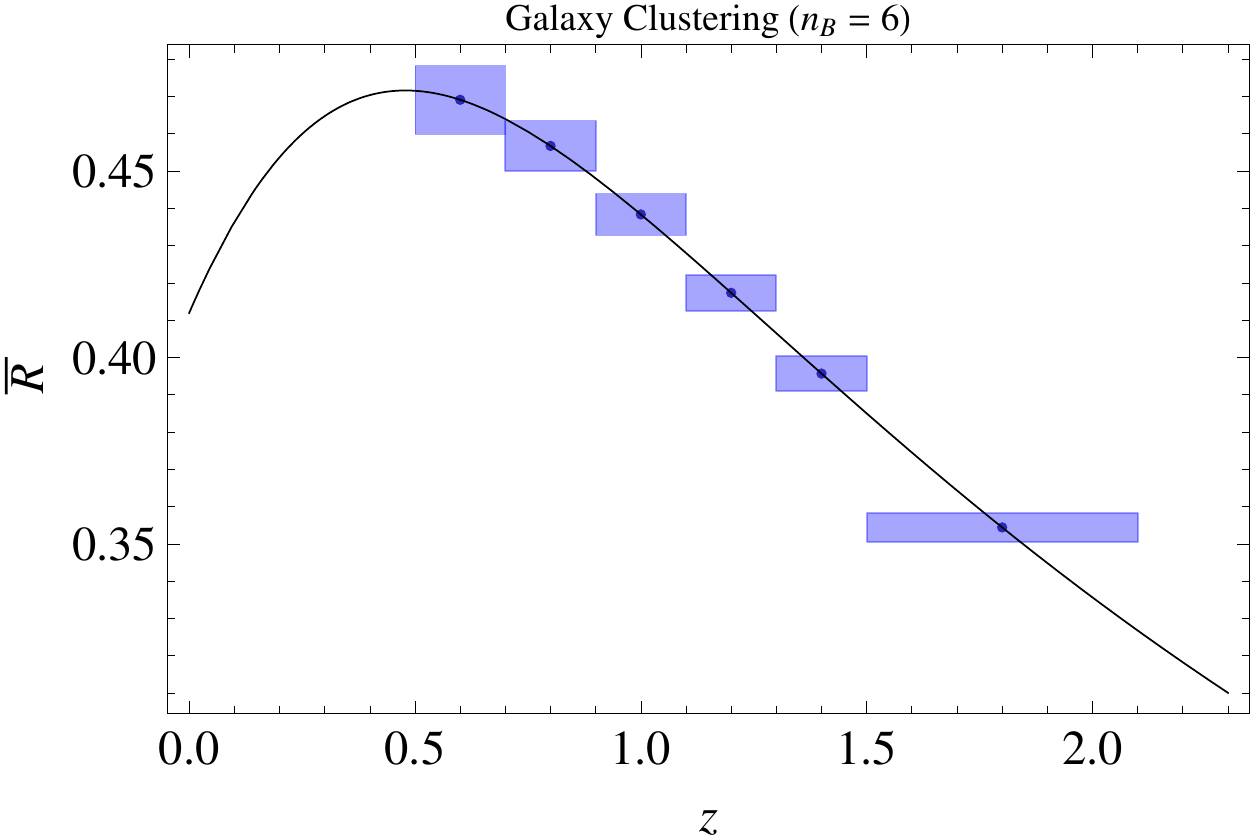} 
\includegraphics[width=0.32\linewidth]{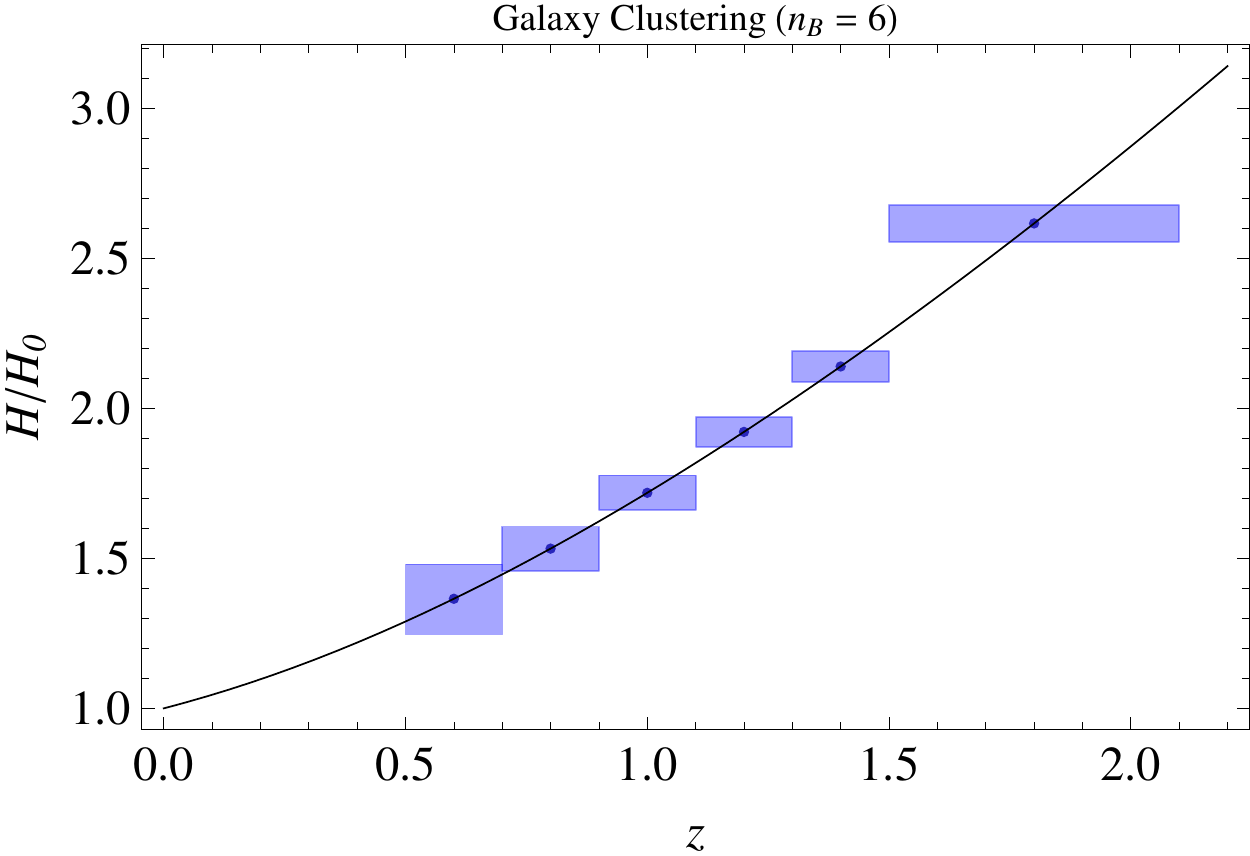} 
\caption{\label{fig1}Errors on $\bar{A}$, $\bar{R}$ and $E$ from Galaxy Clustering in the $z$-binning case.}
\end{figure}

The errors in the set of parameters
$p_{\alpha}^{\text{\tiny{GC}}}$ are taken from the
square root of the diagonal elements of the inverted Fisher matrix, i.e. the
errors are marginalized over all other parameters.  In
Table~\ref{tab:pamsGC1} we present the fiducial values for $\bar{A}$,
$\bar{R}$ and $E$ evaluated at the center of the bins ($\bar{z}_a$), and the
respective errors, and in Fig.~\ref{fig1} we plot their
fiducial values and errors.  

\begingroup \squeezetable 
\begin{table}[b]
\begin{tabular}{ccccccccccc}
\hline \hline
$\bar{z}$  & $\bar{n}(\bar{z})\times10^{-3}$  
& $\bar{A}$  & $\Delta\bar{A}$  
& $\Delta\bar{A}(\%)$  & $\bar{R}$  
& $\Delta\bar{R}$  & $\Delta\bar{R}(\%)$  
& $E$  & $\Delta E$  & $\Delta E(\%)$\tabularnewline
\hline 
0.6  & 3.56  & 0.612  & 0.0022  & 0.37  & 0.469  & 0.0092  & 2.0  & 1.37  & 0.12  & 8.5 \tabularnewline
0.8  & 2.42  & 0.558  & 0.0017  & 0.3  & 0.457  & 0.0068  & 1.5  & 1.53  & 0.073  & 4.8 \tabularnewline
1.0  & 1.81  & 0.511  & 0.0015 & 0.29  & 0.438  & 0.0056  & 1.3  & 1.72  & 0.058  & 3.4\tabularnewline
1.2  & 1.44  & 0.47  & 0.0014  & 0.29  & 0.417  & 0.0049  & 1.2  & 1.92  & 0.05  & 2.6 \tabularnewline
1.4  & 0.99  & 0.434  & 0.0015  & 0.35  & 0.396  & 0.0047  & 1.2  & 2.14  & 0.051  & 2.4 \tabularnewline
1.8  & 0.33  & 0.377  & 0.0018  & 0.47  & 0.354  & 0.0039  & 1.1  & 2.62  & 0.061  & 2.3 \tabularnewline
\hline \hline
\end{tabular}\caption{\label{tab:pamsGC1}Fiducial values and errors for $\bar{A}$, $\bar{R}$
and $E$ using six redshift bins. Units of galaxy number densities are ($h/$Mpc)$^{3}$.}
\end{table}
\endgroup

If we use a redshift dependent bias $b(z)$ (for
instance taking the values from the Euclid specifications, see
\cite{2011arXiv1110.3193L,Orsi:2009mj}), we get only slight deviations from the
errors found for the previous case, as we can see in
Table~\ref{tab:pamsGC1bias}.  Thus, our choice of a bias equal to unity does
not impact the Fisher errors significantly.

\begingroup \squeezetable 
\begin{table}[t]
\begin{tabular}{ccccccccccc}
\hline \hline
$\bar{z}$  &  $\bar{A}$  & $\Delta\bar{A}$  & $\Delta\bar{A}(\%)$  & $\bar{R}$  
& $\Delta\bar{R}$  & $\Delta\bar{R}(\%)$  & $E$  & $\Delta E$  & $\Delta E(\%)$\tabularnewline
\hline 
0.6  &  0.645  & 0.0023  & 0.36  & 0.469  & 0.0094  & 2.  & 1.37  & 0.12  & 8.8 \tabularnewline
0.8  &  0.628  & 0.0018  & 0.28  & 0.457  & 0.0072  & 1.6  & 1.53  & 0.078  & 5.1 \tabularnewline
1.0  &  0.575  & 0.0015  & 0.26  & 0.438  & 0.0059  & 1.3  & 1.72  & 0.06  & 3.5 \tabularnewline
1.2  &  0.584  & 0.0014  & 0.24  & 0.417  & 0.0052  & 1.2  & 1.92  & 0.053  & 2.7 \tabularnewline
1.4  &  0.561  & 0.0015  & 0.27  & 0.396  & 0.005  & 1.3  & 2.14  & 0.053  & 2.5 \tabularnewline
1.8  &  0.561  & 0.0015  & 0.26  & 0.354  & 0.0038  & 1.1  & 2.62  & 0.056  & 2.1 \tabularnewline
\hline \hline
\end{tabular}\caption{\label{tab:pamsGC1bias}Fiducial values and errors for $\bar{A}$, $\bar{R}$
and $E$ using six bins, considering a redshift dependent bias.}
\end{table}
\endgroup

\subsection{$k$ binning}

\label{subsec:gck}

For the third and fourth model we also need a binning in $k$-space. Since ultimately we would like
to obtain error estimates on three functions, $h_2,h_4,h_5$, we will need a minimum of three $k$-bins, 
which is the choice we make here. 
We denote with latin indexes $a,b,c...$ the $z$ bins and with indexes $i,j,k...$ the $k$ bins.
So for the first $z$-bin we have as parameters $s_1=\{\bar{A}_{11},\bar{R}_{12},E_1   \}$, for the second
$s_2=\{\bar{A}_{21},\bar{R}_{22},E_2   \}$, and so forth, with
 $\bar{A}_{ai}=\bar{A}(\bar{z}_a,\bar{k}_{i})$, $\bar{R}_{ai}=\bar{R}(\bar{z}_a,\bar{k}_{i})$, 
and $E_a = E(\bar{z}_a)$, where $\bar{k}_{i}$ denote the centers of the $k$-bins.
The set of parameters is therefore $p_{\alpha}^{\text{\tiny{GC}}} = \{s_1,s_2,...\}$.
The Fisher matrix integration over $k$ is split
into three $k$-ranges between
$k_{\text{max}}$ and $k_{\text{min}}$ which we choose so that $\Delta\log k=\rm const$. The Fisher matrix 
becomes then
\begin{equation}\label{Fishgck}
F_{\alpha\beta}^{\text{GC}}=\frac{1}{8\pi^{2}}\int_{-1}^{1}
d\mu\int_{\Delta k}k^{2} V_{\text{eff}} D_{\alpha}D_{\beta}\,{\mathrm d k} \, ,
\end{equation}
with $\Delta k$ denoting the respective range of the integration.
Denoting
the entry $F_{\bar{A}\bar{R}}$ as $\bar{A}\bar{R}$ , and so on, we can represent the structure of the matrix for every redshift bin
as follows: 
\begin{equation}
\left(\begin{array}{ccccccc}
\bar{A}_{1}\bar{A}_{1} & \bar{A}_{1}\bar{R}_{1} & 0 & 0 & 0 & 0 & \bar{A}_{1}E\\
\bar{R}_{1}\bar{A}_{1} & \bar{R}_{1}\bar{R}_{1} & 0 & 0 & 0 & 0 & \bar{R}_{1}E\\
0 & 0 & \bar{A}_{2}\bar{A}_{2} & \bar{A}_{2}\bar{R}_{2} & 0 & 0 & \bar{A}_{2}E\\
0 & 0 & \bar{R}_{2}\bar{A}_{2} & \bar{R}_{2}\bar{R}_{2} & 0 & 0 & \bar{R}_{2}E\\
0 & 0 & 0 & 0 & \bar{A}_{3}\bar{A}_{3} & \bar{A}_{3}\bar{R}_{3} & \bar{A}_{3}E\\
0 & 0 & 0 & 0 & \bar{R}_{3}\bar{A}_{3} & \bar{R}_{3}\bar{R}_{3} & \bar{R}_{3}E\\
E\bar{A}_{1} & E\bar{R}_{1} & E\bar{A}_{2} & E\bar{R}_{2} & E\bar{A}_{3} & E\bar{R}_{3} & EE
\end{array}\right),
\end{equation}

  In Table~\ref{fid3} we display the values for the integration limits at 
every redshift (the $k$-bins borders), and in Table~\ref{tab:errorsARE}
 we present the errors for all $(z,k)$-bins. Notice that the errors on $E$ are not affected
by the $k$-binning, as $E$ does not depend on $k$.

\begingroup \squeezetable 
\begin{table}[b]
\begin{tabular}{ccccccccccc}
\hline \hline
$\bar{z}$  & $i$  & $\bar{A}$  & $\Delta\bar{A}$  & $\Delta\bar{A}(\%)$  
& $\bar{R}$  & $\Delta\bar{R}$  & $\Delta\bar{R}(\%)$  & $E$ 
 & $\Delta E$  & $\Delta E(\%)$\tabularnewline
\hline \multirow{3}{*}{0.6}  & 1  & \multirow{3}{*}{0.612}  & 0.025  & 4.  & \multirow{3}{*}{0.469}  & 0.07  & 15.  & \multirow{3}{*}{1.37}  & \multirow{3}{*}{0.11}  & \multirow{3}{*}{8.4} \tabularnewline
 & 2  &  & 0.0058  & 0.94  &  & 0.017  & 3.6  &  &  & \tabularnewline
 & 3  &  & 0.0023  & 0.38  &  & 0.0097  & 2.1  &  &  & \tabularnewline
\hline 
\multirow{3}{*}{0.8}  & 1  & \multirow{3}{*}{0.558}  & 0.018  & 3.2 & \multirow{3}{*}{0.457}  & 0.05  & 11 & \multirow{3}{*}{1.53}  & \multirow{3}{*}{0.074}  & \multirow{3}{*}{4.8} \tabularnewline
 & 2 &  & 0.0039  & 0.71  &  & 0.012  & 2.6  &  &  & \tabularnewline
 & 3  &  & 0.0018  & 0.32  &  & 0.0074  & 1.6  &  &  & \tabularnewline
\hline 
\multirow{3}{*}{1.0}  & 1  & \multirow{3}{*}{0.511}  & 0.014  & 2.7  & \multirow{3}{*}{0.438}  & 0.039  & 8.9  & \multirow{3}{*}{1.72}  & \multirow{3}{*}{0.058}  & \multirow{3}{*}{3.4} \tabularnewline
 & 2  &  & 0.003  & 0.59  &  & 0.0089 & 2.  &  &  & \tabularnewline
 & 3  &  & 0.0016  & 0.31  &  & 0.0062  & 1.4  &  &  & \tabularnewline
\hline 
\multirow{3}{*}{1.2}  & 1  & \multirow{3}{*}{0.47}  & 0.011  & 2.4  & \multirow{3}{*}{0.417}  & 0.032  & 7.7  & \multirow{3}{*}{1.92}  & \multirow{3}{*}{0.051}  & \multirow{3}{*}{2.6} \tabularnewline
 & 2  &  & 0.0025  & 0.54  &  & 0.0072  & 1.7  &  &  & \tabularnewline
 & 3  &  & 0.0015  & 0.32  &  & 0.0055  & 1.3  &  &  & \tabularnewline
\hline 
\multirow{3}{*}{1.4}  & 1  & \multirow{3}{*}{0.434}  & 0.01  & 2.3 & \multirow{3}{*}{0.396}  & 0.028  & 7.  & \multirow{3}{*}{2.14}  & \multirow{3}{*}{0.052}  & \multirow{3}{*}{2.4} \tabularnewline
 & 1  &  & 0.0024  & 0.55  &  & 0.0065  & 1.6  &  &  & \tabularnewline
 & 3  &  & 0.0018  & 0.41  &  & 0.0057  & 1.4  &  &  & \tabularnewline
\hline 
\multirow{3}{*}{1.8}  & 1 & \multirow{3}{*}{0.377}  & 0.0063  & 1.7  & \multirow{3}{*}{0.354}  & 0.015  & 4.3  & \multirow{3}{*}{2.62}  & \multirow{3}{*}{0.059}  & \multirow{3}{*}{2.3} \tabularnewline
 & 2 &  & 0.0022  & 0.58  &  & 0.0047  & 1.3  &  &  & \tabularnewline
 & 3  &  & 0.0024 & 0.64  &  & 0.0061  & 1.7  &  &  & \tabularnewline
\hline \hline
\end{tabular}\caption{\label{tab:errorsARE} Relative errors for $\bar{A}$, $\bar{R}$ and $E$
at every redshift and every $k$-bin (labeled with the index $i$). Since fiducial values for $\bar{A}$, $\bar{R}$
and $E$ are independent of $k$, these are the same for the three
$k$-bins. }
\end{table}

\endgroup


\section{Weak lensing\label{sec:wl}}

We move now to estimating the Fisher matrix for a future weak lensing
survey. The lensing convergence power spectrum from a survey divided
into several redshift bins (same binning as in Sec.~\ref{sec:gc})
can be written as \cite{Hu:1999ek} 
\begin{equation}
P_{ij}(\ell)=H_{0}\int_{0}^{\infty}p_{ij}(z,\ell){\mathrm d z} \approx H_{0}\sum_{a}\frac{\Delta
 z_{a}}{E_{a}}K_{i}K_{j}\bar{L}^{2}\delta_{\text{t},0}^{2}\left(\bar
 z_{a},k(\ell,\bar z_{a})\right) \, , \label{eq:ConvInt}
\end{equation}
with the integrand
\begin{equation}
p_{ij}(z,\ell)=\frac{K_{i}(z)K_{j}(z)}{E(z)}\
\bar{L}(z)^{2}\delta_{\text{t},0}^{2}\left(z,k(\ell,z)\right),
\label{eq:integrand}
\end{equation}
 where 
\begin{equation}
k(\ell,z)=\frac{\ell}{\pi r(z)}  \label{eq:zellk}
\quad \mathrm{and} \quad
K_{i}(z)=\frac{3}{2}(1+z)W_{i}(z) \, ,
\end{equation}
and $W_{i}(z)$ is the weak lensing window function for the $i$-th
bin 
\begin{equation}
W_{i}(z)=H_{0}\int_{z}^{\infty}\left(1-\frac{\hat{r}(z)}{\hat{r}(\tilde{z})}\right)n_{i}(\tilde{z}) \, {\mathrm d \tilde{z}} \, .
\end{equation}
 Here, $n_{i}(z)$ equals the galaxy density $n(z)$ if $z$ lies inside the $i$-th
 redshift bin and zero otherwise.
Note that 
\begin{equation}
n_{i}(z){{\rm d}z=\frac{n_{i}(r(z))}{H(z)}{{\rm d}r}}\,.
\end{equation}
The overall galaxy density is modeled as 
\begin{equation}
n(z)\propto z^{a}\exp(-(z/z_{p})^{b}).\label{eq:wlnz}
\end{equation}
 We take $a=2$, $b=3/2$ and choose $z_{p}$ such that the median of the
distribution is at $z=0.9$, i.e. $z_{p}=0.9/1.412=0.6374$
\cite{Ma2006,2011arXiv1110.3193L}.
The $n_{i}(z)$ (which are not to be confused with the $\bar{n}(z)$
from Galaxy Clustering) are then smoothed with a Gaussian to account
for the photometric redshift error (see \cite{Ma2006}) and normalized
such that $\int n_{i}(z){{\rm d}z=1}$.
Following the Euclid specifications,
we set the survey sky fraction $f_{{\rm sky}}=0.375$ and the photometric redshift error to
 $\delta z = 0.05(1+z)$. 

Including the noise due to intrinsic galaxy ellipticities we have
\begin{equation}
C_{ij}=P_{ij}+\gamma_{{\rm int}}^{2}\hat{n}_{i}^{-1}\delta_{ij},
\end{equation}
 with the intrinsic ellipticity $\gamma_{{\rm int}}=0.22$ and the
number of all galaxies per steradian in the $i$-th bin, $\hat{n}_{i}$,
which can be written as 
\begin{equation}
\hat{n}_{i}=n_{\theta}\frac{\int_{z_{i}}^{z_{i+1}}n(z)\mathrm{d}z}{\int_{0}^{\infty}n(z)\mathrm{d}z},
\end{equation}
where $n_{\theta}$ is the areal galaxy density, an important parameter
that defines the quality of a weak lensing experiment. We set it to
$n_{\theta}=35$ galaxies per square arc minute
\cite{2011arXiv1110.3193L}.

For a weak lensing survey that covers a fraction of the sky $f_{{\rm sky}}$,
the Fisher matrix is a sum over $\ell$ bins of size $\Delta\ell$
\citep{Eisenstein1999a}
\begin{equation}
F_{\alpha\beta}^{\text{WL}}=f_{{\rm
sky}}\sum_{\ell}\Delta\ell \frac{(2\ell+1)}{2}\frac{\partial
P_{ij}}{\partial p_{\alpha}}C_{jm}^{-1}\frac{\partial P_{mn}}{\partial
p_{\beta}}C_{ni}^{-1},
\label{eq:wlfm}
\end{equation} 
 and now the parameters are
$p_{\alpha}=\{\bar{L}(\bar z_{1}),E(\bar z_{1}),\dots\}$.  Here, $\ell$ is being summed 
from 5 to $\ell_{\rm max}$ with
$\Delta\log\ell=0.1$, where $\ell_{\rm max}$ corresponds to the value listed
in Table~\ref{tab:EL5000} for the redshift bin $a$ or $b$ --- whichever is
smaller.

\begin{figure}[!t]
 \includegraphics[width=0.46\linewidth]{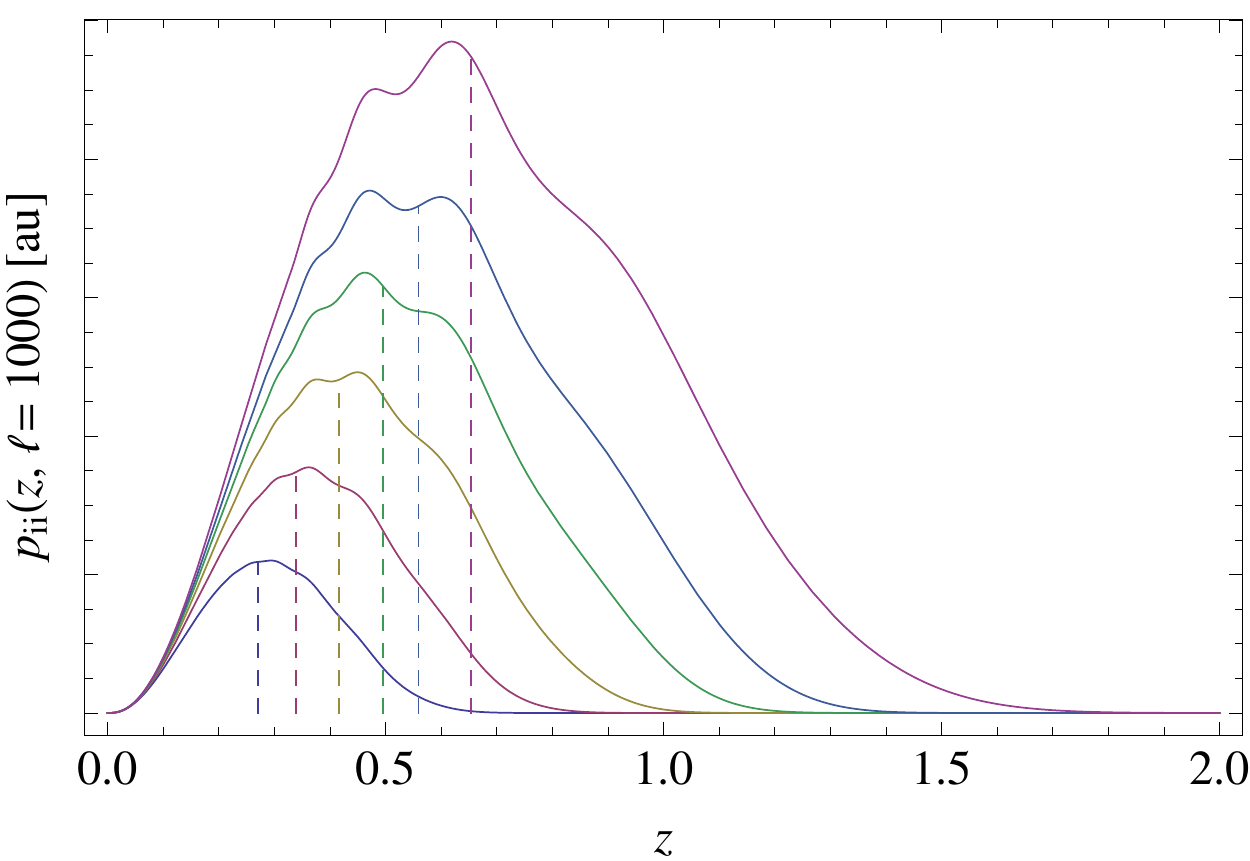} 
 \caption{ The  integrand of Eq.~(\ref{eq:ConvInt}). The curves from left to
 right correspond to $p_{ii}(z,\ell=1000)$, where $i=1,\dots,6$.
The contribution to the lensing signal is very broad in redshift and
peaks at relatively low $z$ even for the high-redshift bins. The median
redshift for each curve is indicated by dashed lines. We give
the median redshift for the lensing contribution in Table~\ref{tab:EL5000}.}
\label{fig:integrands} 
\end{figure}

The value $\ell_{\rm max}$ is derived as follows.
We start with the relationship
\begin{equation} 
\frac{\ell}{\pi r(z_{{\rm med}}(\ell,a))}=k,\label{eq:ellmax} 
\end{equation} 
where $z_{{\rm med}}(\ell,a)$ is the median with respect to $z$ of
$p_{aa}(z,\ell)$, which is defined in Eq.~(\ref{eq:integrand}). For a given
wave number $k$ and a redshift bin $a$, we can solve for $\ell$.  To find
$\ell_{\rm max}$ we use the following method:

We begin with $z_{\rm med} = 1$, compute the $k_{\rm max}$ for this
redshift as before by imposing $\sigma^2(R)=0.35$, solve
Eq.~(\ref{eq:ellmax}) for $\ell$, and compute $z_{{\rm med}}(\ell,a)$. We
repeat this step until the value for $z_{\rm med}$ converges with an
accuracy of approximately 1\%.
A list of the values for $\ell_{{\rm max}}$ as well as $z_{{\rm med}}$ used
in each redshift bin can be found in Table~\ref{tab:EL5000}. The integrands
along with their median value are depicted in Fig.~\ref{fig:integrands}.

\begingroup \squeezetable 
\begin{table}[b]
\begin{tabular}{ccccccccc}
\hline \hline
$\bar{z}$  & $\ell_{{\rm max}}$  & $z_{{\rm med}}$  
& $\bar{L}$  & $\Delta\bar{L}$  
& $\Delta\bar{L}(\%)$  & $E$  & $\Delta E$  & $\Delta E(\%)$ \tabularnewline
\hline 
 0.6 & 311  & 0.26 & 0.342 & 0.0044 & 1.3 & 1.37 & 0.0062 & 0.46 \\
 0.8 & 385  & 0.31 & 0.311 & 0.0044 & 1.4 & 1.53 & 0.0069 & 0.45 \\
 1.0 & 515  & 0.40 & 0.285 & 0.0059 & 2.1 & 1.72 & 0.017 & 0.96 \\
 1.2 & 609  & 0.45 & 0.262 & 0.0059 & 2.3 & 1.92 & 0.029 & 1.5 \\
 1.4 & 760  & 0.54 & 0.242 & 0.014 & 5.7 & 2.14 & 0.029 & 1.4 \\
 1.8 & 959  & 0.64 & 0.210 & 0.035 & 16 & 2.62 & 0.077 & 3.0 \\

\hline \hline
\end{tabular}\caption{Errors on $E$ and $\bar{L}$ from
weak lensing only (with six redshift bins) and a list of the value
$\ell_{{\rm max}}$ 
used at each redshift together with the corresponding $z_{{\rm med}}$
value.}
\label{tab:EL5000} 
\end{table}
\endgroup

To find the derivatives needed in Eq.~(\ref{eq:wlfm}), we divide the
integral in Eq.~(\ref{eq:ConvInt}) into $n_B$ integrals that each cover one
redshift bin. We could assume that $\bar L(z)$ is constant across any
redshift bin to get an approximate expression for the integral that depends
on $\bar L$ in an analytical way, but the discrepancy between the
actual integral and the approximate integral (and consequently the
discrepancy of the derivative) can be up to a factor of 2, which may not be
sufficient. Assuming that the integrand is linear in $z$ gives the same
result (when using only the center of the bin as the sampling point), so the
issue arises when the curvature of the integrand becomes large. 

As a solution, we take the actual value of the integral and simply assume
that it depends quadratically on $\bar L(\bar z_a)$, such that the derivative
can be written as
\begin{equation}
\frac{\partial P_{ij}(\ell)}{\partial\bar{L}(\bar z_a)}=
\frac{2}{\bar L(\bar z_a)} \int_{z_a}^{z_{a+1}}
p_{ij}(z,\ell) \mathrm dz.
\label{eq:PijL}
\end{equation} 

Since $E$ appears in the comoving distance, it is more complicated for the
derivatives of $P_{ij}$ with respect to $E(\bar z_a)$. We substitute the
regular definition of $E$ by an interpolating function that goes smoothly
through all points $(\bar z_a,E(\bar z_a))$ and $(0,1)$. Instead of
depending on $\Omega_{m}$ it now depends on the values of all $E(\bar z_a)$,
and so do all functions that depend on $E$, in particular the comoving
distance and consequently the window functions $K_{i}(z)$. The derivatives
are then obtained by varying the fiducial values of $E(\bar z_a)$ while
keeping $L=\bar{L} \delta_{\text{t,0}}$ fixed so that we again do not include the
derivative of $\delta_{\text{t,0}}^2$ with respect to $k$.

It is instructive to consider the error on the spectrum itself for
a particular pair $ij$. If we take as parameters $p_{\alpha}=P_{ij}$
we have a variance
\begin{equation}
\sigma^{-2}=f_{{\rm sky}}\sum_{\ell}\Delta\ell\frac{(2\ell+1)}{2}C_{ij}^{-1}C_{ij}^{-1},
\end{equation}
 (no sum over $ij$) and neglecting the noise (appropriate for $\ell<500$)
i.e. putting $C_{ij}=P_{ij}$, this becomes, in a small range of $\ell$
from $\ell_{\text{min}}$ to $\ell_{\text{max}}$
so that we can approximate $P_{ij}$ with a constant,
\begin{equation}
\sigma^{-2}P_{ij}P_{ij}=f_{{\rm sky}}\sum_{\ell}\Delta\ell\frac{(2\ell+1)}{2}=f_{{\rm sky}}\frac{\ell_{\text{max}}^{2}-\ell_{\text{min}}^{2}}{2},
\end{equation}
 (for $\ell_{\text{max,min}}\gg1$). If $\ell_{\text{min}}$ is much smaller than $\ell_{\text{max}}$ this gives a relative error for every
$ij$
\begin{equation}
\frac{\sigma}{P_{ij}}=\ell_{\text{max}}^{-1}\left(\frac{f_{{\rm sky}}}{2}\right)^{-1/2}\approx2.3\, \ell_{\text{max}}^{-1},
\end{equation}
 so that for $\ell_{\text{max}}=300$ we should get a minimum relative
error of $0.6\%$, which is indeed of the same order as our result.
The error increases if we include the noise and a non-negligible $\ell_{\text{min}}$.

\begin{figure}[!t]
\centering
 \includegraphics[width=0.46\linewidth]{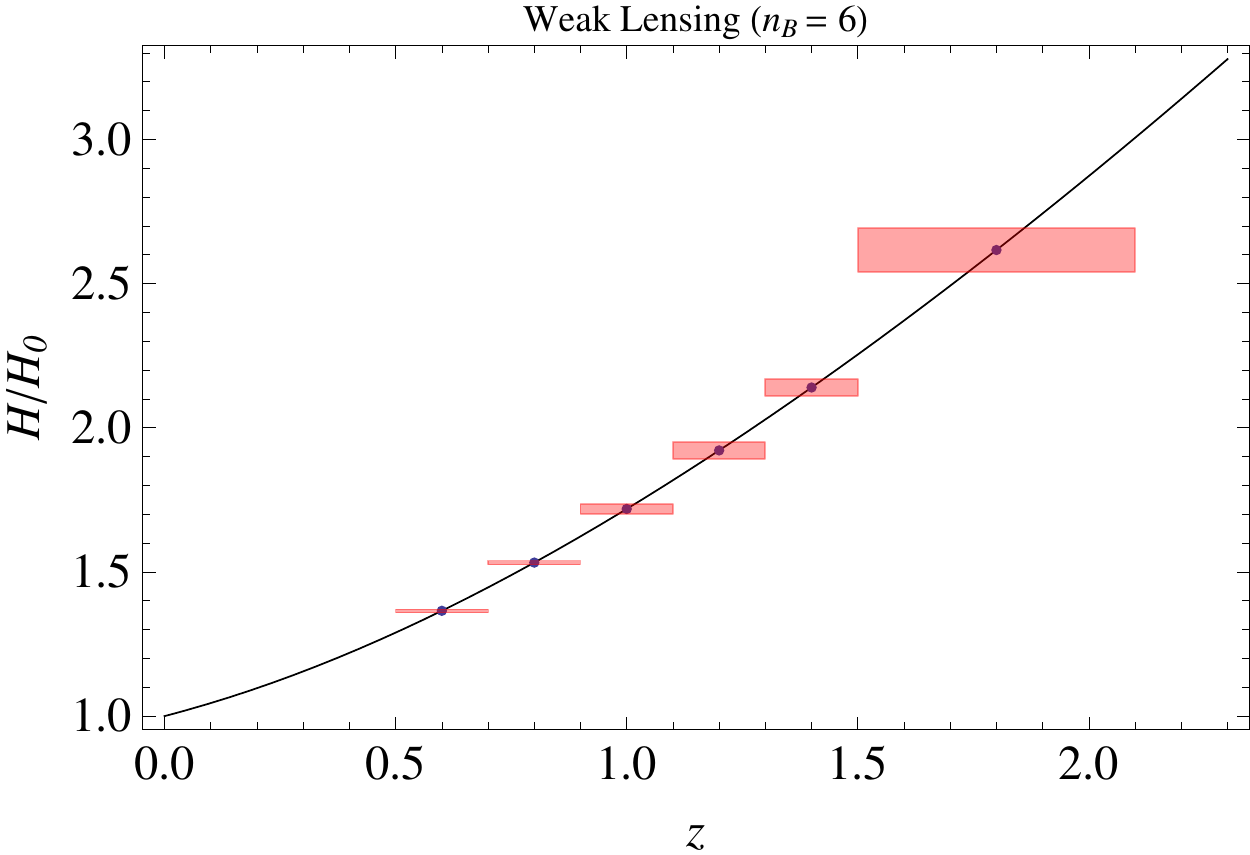} 
 \includegraphics[width=0.46\linewidth]{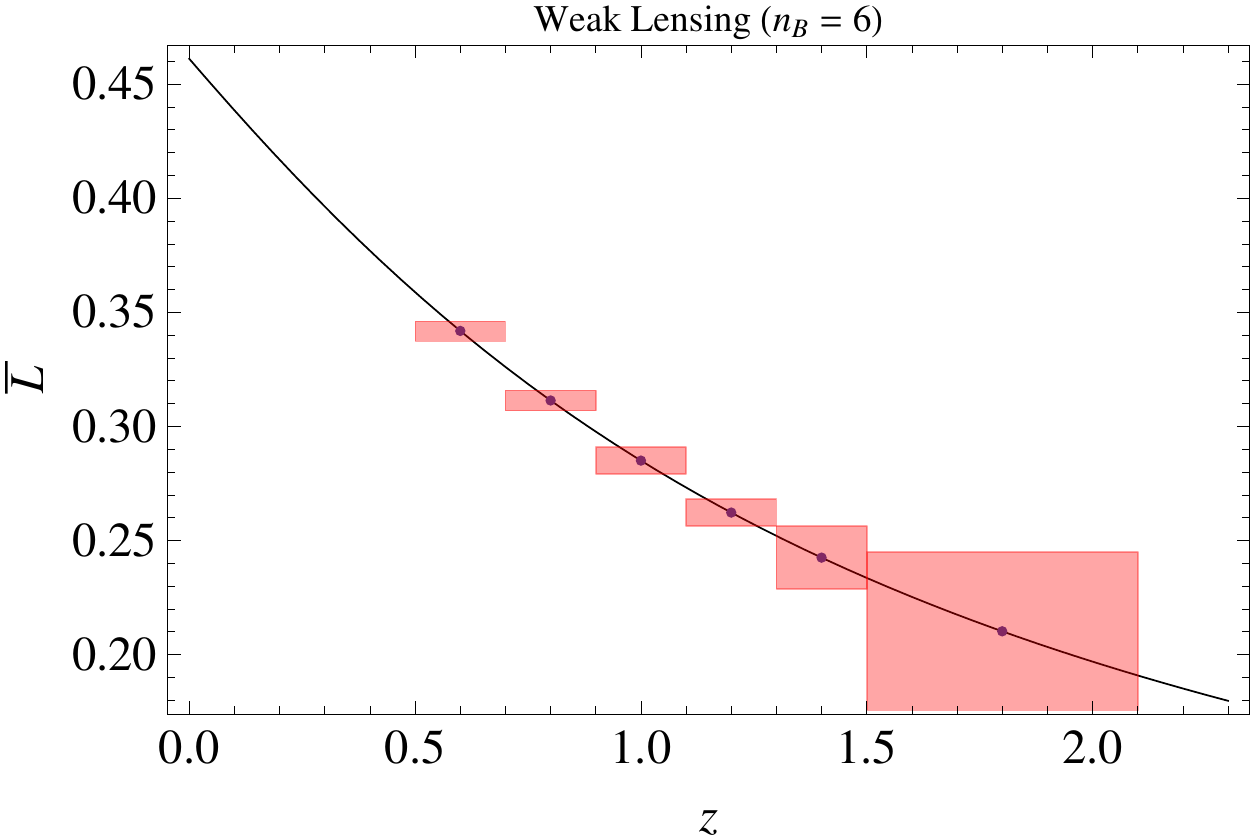} 
\caption{\label{ErrorEWL}Errors on $E(\bar z_a)$ (left) and $\bar{L}(\bar
z_a)$ (right) from weak lensing.}
\end{figure}

The resulting uncertainties on $E(\bar z_a)$ and $\bar{L}(\bar z_a)$
can be found in Table~\ref{tab:EL5000}; they are visualized
in Fig.~\ref{ErrorEWL}. 

\subsection{$k$ binning}

\begingroup \squeezetable 
\begin{table}[b]
\begin{tabular}{ccccc}
\hline \hline
$\bar{z}$  & $\ell_{0}$  & $\ell_{1}$  & $\ell_{2}$  & $\ell_{3}$ \tabularnewline
\hline 
0.6 & 6.3 & 39  & 120 & 410 \tabularnewline
0.8 & 7.9 & 45  & 190 & 610 \tabularnewline
1.0 & 9.4 & 66  & 240 & 880 \tabularnewline
1.2 & 11 & 83  & 320 & 1200 \tabularnewline
1.4 & 12 & 97  & 390 & 1500 \tabularnewline
1.8 & 14 & 120 & 550 & 2200 \tabularnewline
\hline \hline
\end{tabular}\caption{Borders of the $\ell$-bins for each redshift bin converted from the
$k$-bins according to Eq.~(\ref{eq:ellmax}).}
\label{tab:ellbins} 
\end{table}
\endgroup

To test the cases three and four of our models for $\eta$, we need
to consider $\bar{L}$ as a function of $k$ (although with the same
fiducial value for all $k$, as the fiducial model is $\Lambda$CDM),
and we divide the full $k$-range again into the same three bins. The observables
are then $\bar{L}_{an}\equiv\bar{L}(\bar z_a,\bar{k}_{n})$,
where $\bar{k}_{n}$ denote the center of the $k$-bins,  with $n=1,2,3$. They are
defined as in Sec.~\ref{sec:gc}, and are given explicitly in Table~\ref{fid3}.
The $k$-bins fix the ranges
for $\ell$ via the relation used in Eq.~(\ref{eq:ellmax}). We label
the center of the $\ell$-bins accordingly as $\ell_{n}$.
See Table~\ref{tab:ellbins} for a list of the $\ell$-bins. The derivatives
needed for the Fisher matrix will be evaluated at the center of these
$\ell$-bins.

They can be computed similarly as in Eq.~(\ref{eq:PijL}). We find
(using Kronecker deltas, no summation): 
\begin{equation}
\frac{\partial
P_{ij}(\ell)}{\partial\bar L(\bar z_a,k_{n})}= \frac{2\delta_{an}}{\bar L(z_a)}
 \int_{z_a}^{z_{a+1}} p_{ij}(z,\ell) \mathrm dz
 \times 
\begin{cases}
1 & {\rm if} \quad \ell_{n-1} < \ell < \ell_{n}\\ 
0 & {\rm else}.
\end{cases}
\end{equation}
The derivatives with respect to $E(\bar z_a)$ are computed the same way as
before. We can then define the parameter vector
$p_\alpha=\{ \bar L_{11},E_1,\bar L_{12}, E_1,\bar L_{13}, E_3, \bar L_{21},E_2,... \}$
and evaluate the Fisher matrix formally as before.
The structure of the Fisher matrix can be schematically represented as follows:
\begin{equation}
\left(\begin{array}{cccc}
\bar{L}_{1}\bar L_1 & 0 & 0 & \bar{L}_{1}E\\
0 & \bar{L}_{2}\bar{L}_{2} & 0 & \bar{L}_{2}E\\
0 & 0 & \bar{L}_{3} \bar{L}_{3} & \bar{L}_{3}E\\
\bar{L}_{1}E & \bar{L}_{2}E & \bar{L}_{3}E & EE
\end{array}\right)
\end{equation}
The uncertainties placed on the observables by weak lensing only can
be found in Table~\ref{tab:LkzWL}.

\begingroup \squeezetable 
\begin{table}[t]
\begin{tabular}{ccccccccccccc}
\hline \hline
$\bar{z}$  & $\bar L_{a1}$  & $\Delta \bar L_{a1}$  & $\Delta \bar L_{a1}$ (\%)  &
$\bar L_{a2}$  & $\Delta \bar L_{a2}$  & $\Delta \bar L_{a2}$(\%)  &
$\bar L_{a3}$  & $\Delta
\bar L_{a3}$  & $\Delta \bar L_{a3}$(\%)  & $E_{a}$  & $\Delta E_{a}$  & $\Delta
E_{a}$(\%) \tabularnewline \hline 
 0.6 & 0.342 & 0.025 & 7.4 & 0.342 & 0.0076 & 2.2 & 0.342 & 0.0050 & 1.5 & 1.37 & 0.0069 & 0.51 \\
 0.8 & 0.311 & 0.025 & 7.9 & 0.311 & 0.0064 & 2.1 & 0.311 & 0.0053 & 1.7 & 1.53 & 0.0074 & 0.48 \\
 1.0 & 0.285 & 0.022 & 7.8 & 0.285 & 0.0074 & 2.6 & 0.285 & 0.0062 & 2.2 & 1.72 & 0.017 & 0.97 \\
 1.2 & 0.262 & 0.024 & 9.1 & 0.262 & 0.0080 & 3.0 & 0.262 & 0.0073 & 2.8 & 1.92 & 0.030 & 1.6 \\
 1.4 & 0.242 & 0.041 & 17. & 0.242 & 0.019 & 7.7 & 0.242 & 0.015 & 6.1 & 2.14 & 0.030 & 1.4 \\
 1.8 & 0.210 & 0.098 & 46. & 0.210 & 0.048 & 23 & 0.210 & 0.037 & 17 & 2.62 & 0.079 & 3.0 \\
  
\hline \hline
\end{tabular}\caption{Errors of $\bar{L}_{ai}$ and $E$ using weak lensing
only with their fiducial values. }
\label{tab:LkzWL} 
\end{table}
\endgroup


\section{Supernovae\label{sec:sn}}

\begingroup \squeezetable
\begin{table}[b]
\begin{tabular}{cccccc}
\hline \hline
$\bar{z}$  & $\sigma_{{\rm data},a}$  & $n_a$  &
$E(\bar{z})$  & $\Delta E$  & $\Delta E(\%)$
\tabularnewline \hline 
0.6  & 0.287  & 46429  & 1.37  & 0.0026  & 0.19 \tabularnewline
0.8  & 0.285  & 25000  & 1.53  & 0.0041  & 0.27 \tabularnewline
1.0  & 0.329  & 16071  & 1.72  & 0.0086  & 0.50 \tabularnewline
1.2  & 0.327  & 7143  & 1.92  & 0.016  & 0.83 \tabularnewline
1.4  & 0.258  & 5357  & 2.14  & 0.028  & 1.3 \tabularnewline
\hline \hline
\end{tabular}\caption{Redshift uncertainties, number of supernovae, fiducial value of $E$ and errors  for each bin. }
\label{tab:sigman-E_SN} 
\end{table}
\endgroup

\begin{figure}[!t]
\centering
 \includegraphics[width=0.46\linewidth]{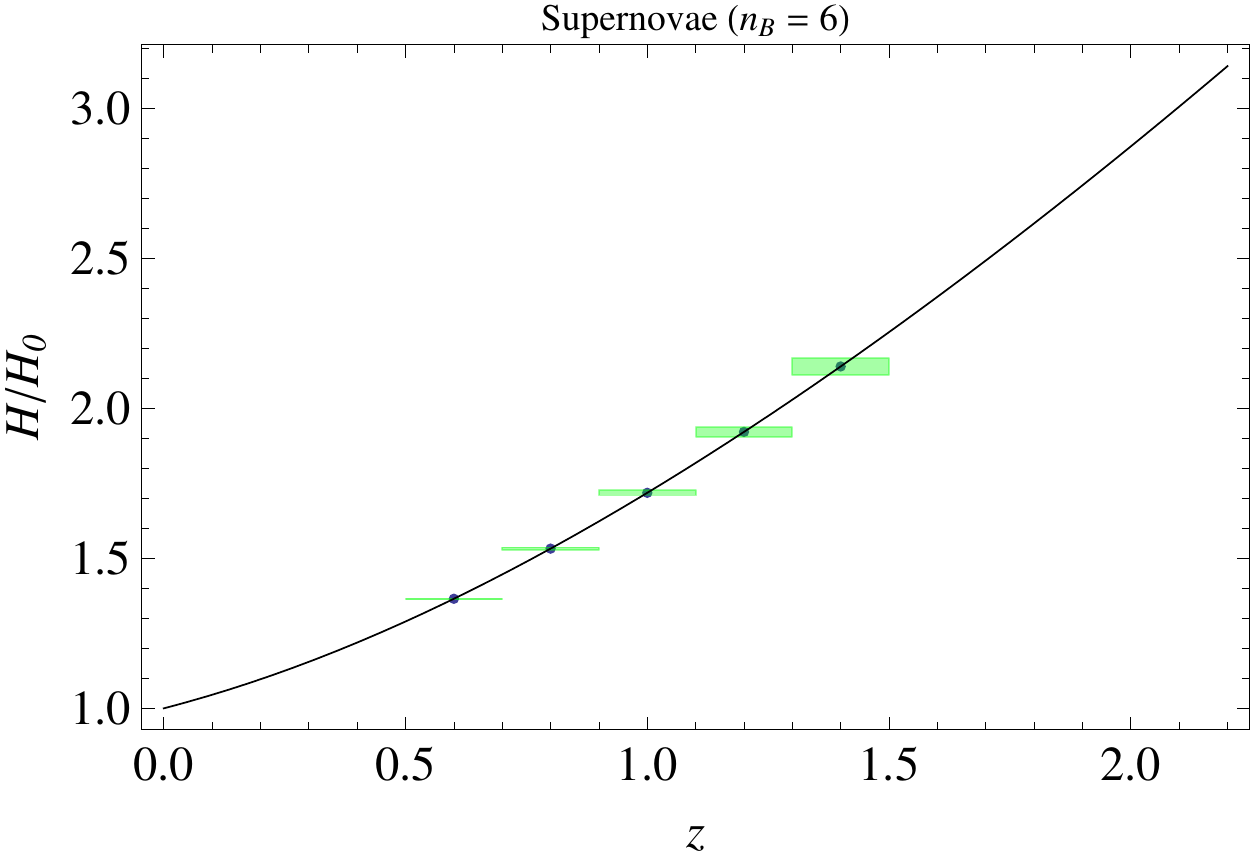}
\caption{\label{figSN1}Errors on $E$ from Supernovae.}
\end{figure}

We consider now the forecasts for a supernovae survey. The likelihood
function for the supernovae after marginalization of the offset is
\cite{Amendola2010} 
\begin{equation}
\mathcal{L}=-\log L=\frac{1}{2}\left(S_{2}-\frac{S_{1}^{2}}{S_{0}}\right),
\end{equation}
where 
\begin{equation}
S_{n}=\sum_{i}\frac{(m_{i}-\mu_{i})^{n}}{\sigma_{i}^{2}},
\end{equation}
 and $\mu_{i}=5\log\hat{d}_{L}$, where $\hat{d}_{L}$ is the dimensionless
luminosity distance, see Eq.~(\ref{eq:dist}). This can be written
as 
\begin{equation}
\mathcal{L}=\frac{1}{2}X_{i}M_{ij}X_{j},
\end{equation}
 where $X_{i}=m_{i}-\mu_{i}$ and 
\begin{equation}
M_{ij}=s_{i}s_{j}\delta_{ij}-\frac{s_{i}^{2}s_{j}^{2}}{S_{0}},
\end{equation}
 (no sum) where $s_{i}=1/\sigma_{i}$. The Fisher matrix can be written
as 
\begin{equation}
F_{\alpha\beta}^{\text{SN}}=\left\langle \frac{\partial\mathcal{L}}{\partial p_{\alpha}}\frac{\partial\mathcal{L}}{\partial p_{\beta}}\right\rangle ,
\end{equation}
 where now the parameters are $p_{\alpha_a}^{\text{\tiny{SN}}}=E(\bar z_{a})$. Similarly to section \ref{sec:gc}
we can write 
\begin{equation}
\hat{d}_{L}(\bar{z}_a)=(1+\bar z_{a})\sum_{b=0}^{b=a}\frac{\Delta z_{b}}{E(\bar z_b)},
\end{equation}
 so that 
\begin{equation}
\frac{\partial\hat{d}_{L}(z_a)}{\partial E(\bar z_{b})}=-\frac{\Delta z_{b}}{E_{b}^{2}}(1+z_{a})\delta_{ab}
\end{equation}
where $\delta_{ab}$ is a Kronecker symbol. The Fisher matrix
for the parameter vector $p_{\alpha}=\{E(z_{a})\}$ with $a$
running over the $z$-bins is then 
\begin{equation}
F_{\alpha\beta}^{\text{SN}}=\left\langle \left(\frac{\partial\mu_{i}}
{\partial p_{\alpha_a}}M_{ij}X_{j}\right)\left(\frac{\partial\mu_{i}}
{\partial p_{\beta_b}}M_{ij}X_{j}\right)\right\rangle =25Y_{i\alpha}M_{ij}Y_{j\beta}.
\end{equation}
 where 
\begin{equation}
Y_{i\alpha}\equiv\frac{\partial\log\hat{d}_{L}(\bar{z}_i)}{\partial
p_{\alpha}}=\frac{1}{\hat{d}_{L}(\bar{z}_i)}\frac{\partial\hat{d}_{L}(\bar{z}_i)}{\partial
E(\bar z_{\alpha})}=-\frac{1}{\hat{d}_{L}(z_{i})}\frac{\Delta
\bar z_{\alpha}}{E_\alpha^{2}}(1+\bar z_{i})\delta_{i\alpha}.\label{eq:Y}
\end{equation}

We have to make a choice to define the redshifts $z_{i}$ and the
uncertainties $\sigma_{i}$ for the supernovae of the simulated future
experiment. We take the Union 2.1 catalog as a reference (580 SNIa
in the range $0<z\lesssim1.5$). We assume that the survey will observe
supernovae in the redshift range $0.5<z<1.5$, and divide that interval
in bins of fixed width $\Delta z=0.2$ just like in Sec.~\ref{sec:gc}, in
order to combine the SN Fisher matrix with the galaxy clustering and the
weak lensing ones.
We assume the total number of observed SN to be about $n_{\text{SN}}=100000$
in that range, as expected for the  LSST survey \cite{Tyson:2003kb}. We further assume that the supernovae of the future
survey will be distributed uniformly in each bin, respecting the proportions
of the data of the catalog Union 2.1 and with the same average magnitude
error. The values of $\sigma_{\text{data},a}$ and $n_{a}$
for the bins centered in $\bar{z}_a$ are summarized in Table~\ref{tab:sigman-E_SN}.

Finally, the corresponding errors on $E$ from supernovae are shown
in Fig.~\ref{figSN1} and in Table~\ref{tab:sigman-E_SN}. In Table~\ref{tab:allerrorsE}
we compare the errors on $E$ from the three different probes with
each other. We notice that the supernova constraints are 
the most stringent ones among the three probes and improve the WL+GC constraints by almost a factor of two.
All this of course assumes that systematic errors
can be kept below statistical errors.


\begin{figure}[t!]
\centering
\includegraphics[width=0.32\linewidth]{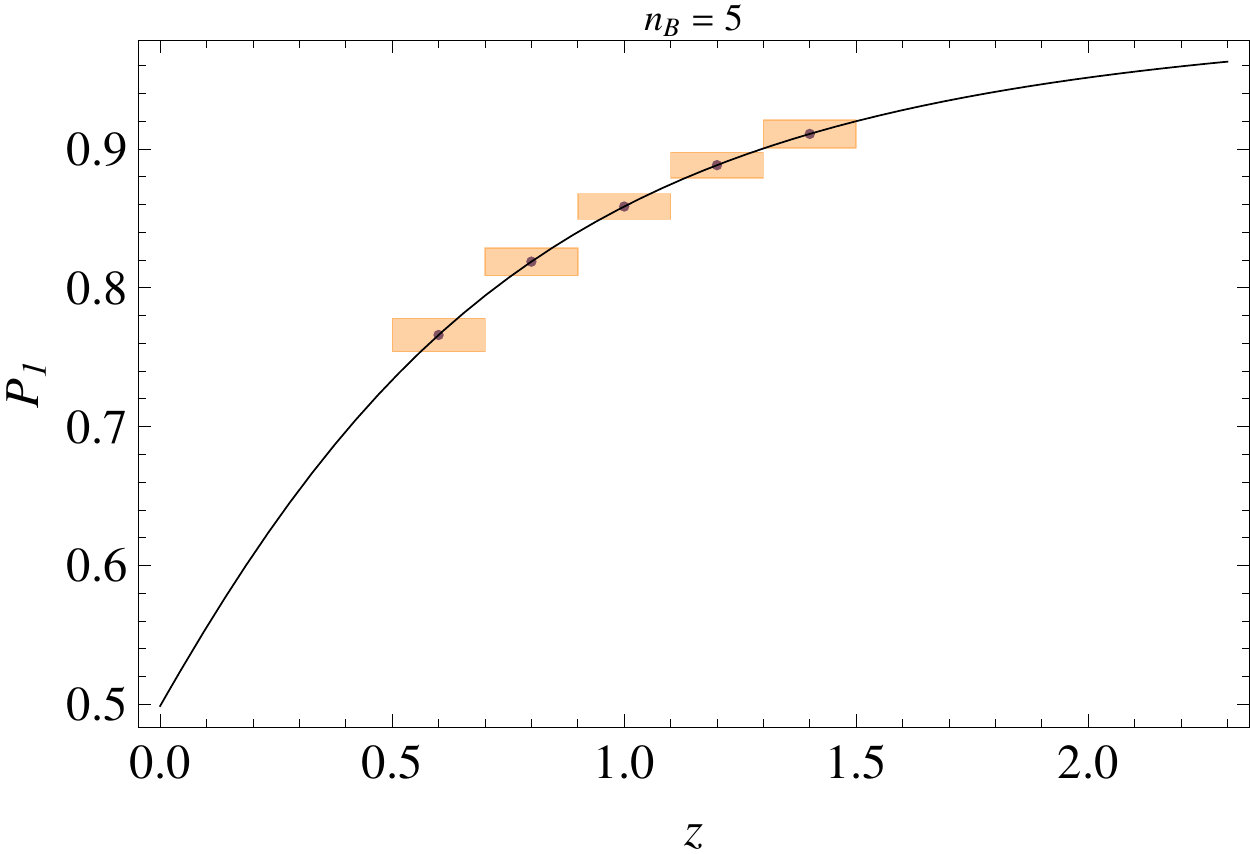} 
\includegraphics[width=0.32\linewidth]{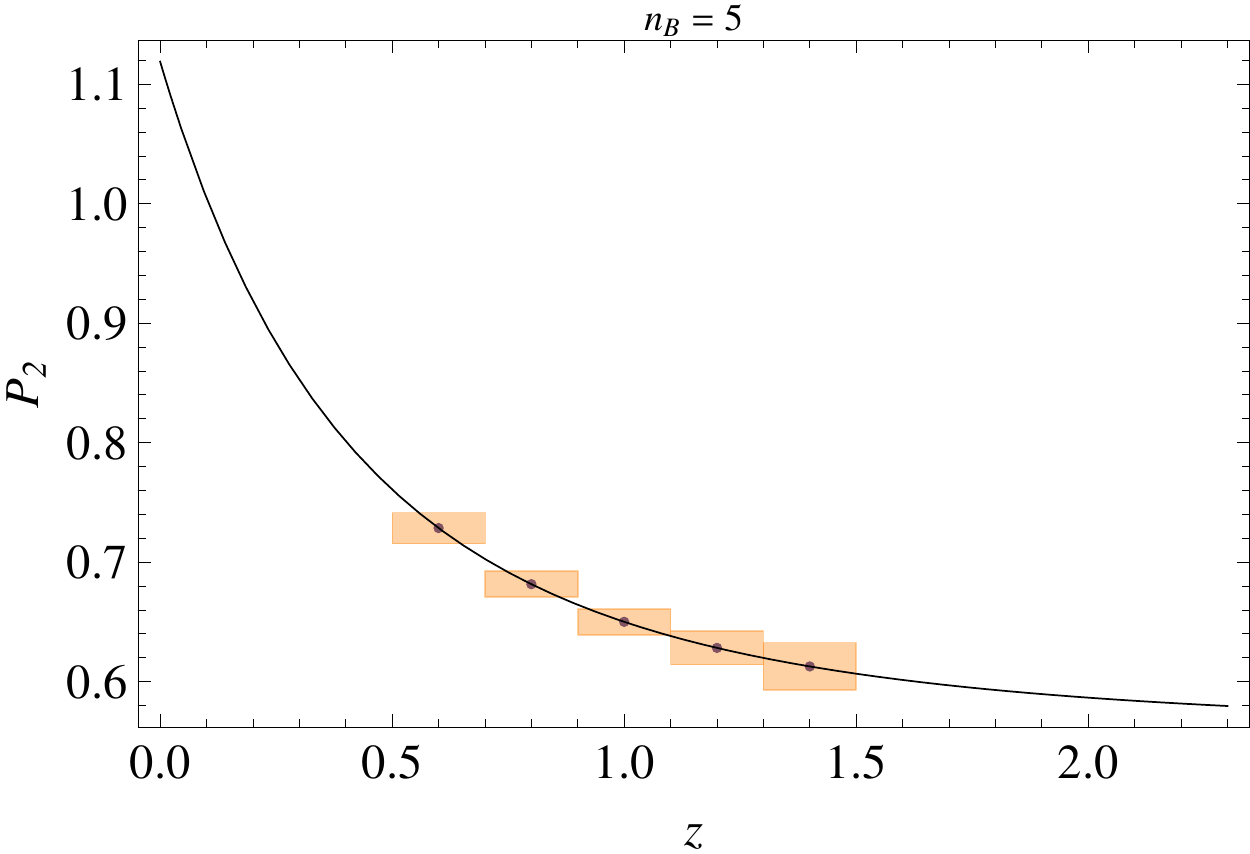} 
\includegraphics[width=0.32\linewidth]{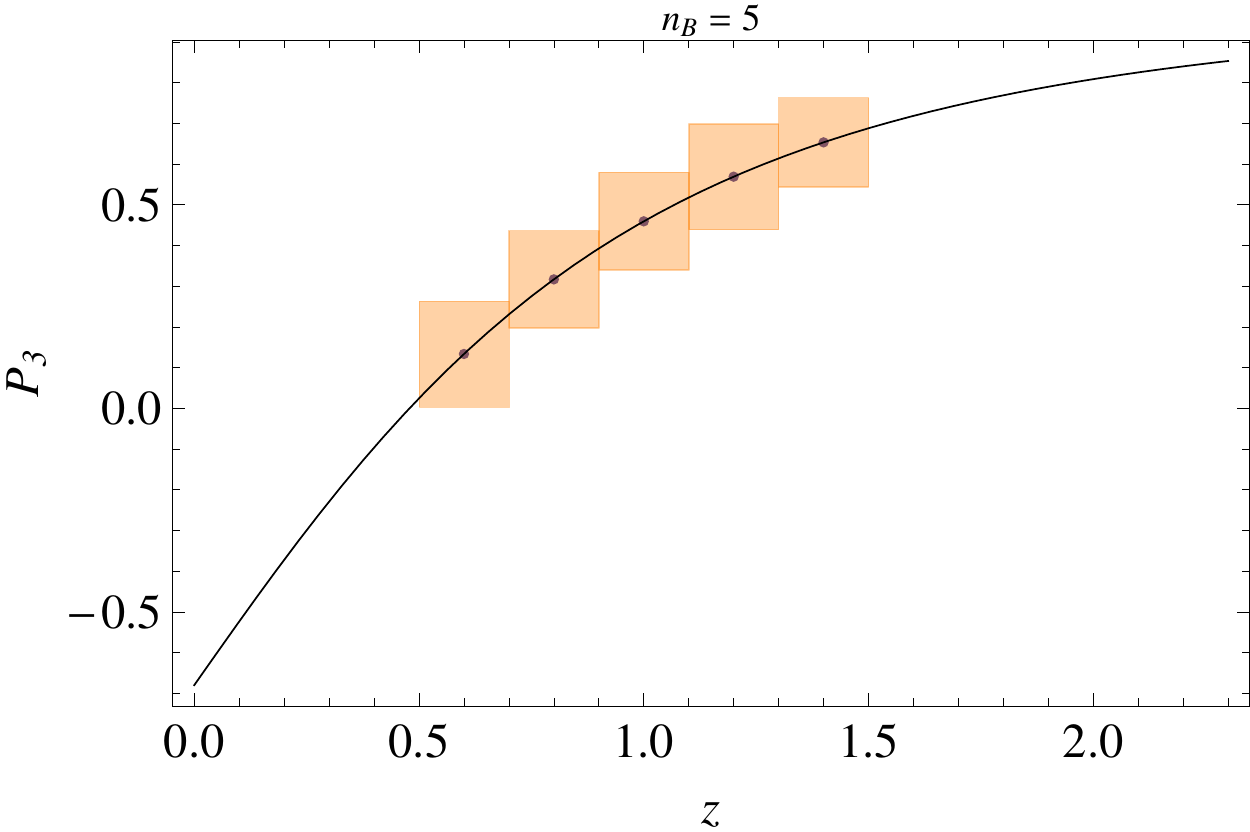} 
\caption{\label{figerrPi}Errors on $P_1$, $P_2$ and $P_3$ in the $z$-varying case.}
\end{figure}

\begingroup 
\squeezetable
\begin{table}[t]
\centering
\begin{tabular}{cccccccccccccccc}
\hline \hline
$\bar{z}$  & $P_{1}$  & $\Delta P_{1}$  & $\Delta P_{1}(\%)$  &
$P_{2}$  &
$\Delta P_{2}$  & $\Delta P_{2}(\%)$  & $P_{3}$  & $\Delta P_{3}$  & $\Delta
P_{3}(\%)$  & $(E'/E)$  & $\Delta E'/E$ & $\Delta E'/E (\%)$ &
$\bar\eta$  & $\Delta\bar\eta$  & $\Delta\bar\eta(\%)$ \tabularnewline
\hline 
 0.6 & 0.766 & 0.012 & 1.6 & 0.729 & 0.013 & 1.8 & 0.134 & 0.13 & 99 & -0.920 & 0.022 & 2.4 & 1 & 0.11 & 11 \\
 0.8 & 0.819 & 0.010 & 1.2 & 0.682 & 0.011 & 1.6 & 0.317 & 0.12 & 38 & -1.04 & 0.046 & 4.4 & 1 & 0.091 & 9.1 \\
 1.0 & 0.859 & 0.0093 & 1.1 & 0.650 & 0.011 & 1.7 & 0.460 & 0.12 & 26 & -1.13 & 0.099 & 8.7 & 1 & 0.090 & 9.0 \\
 1.2 & 0.888 & 0.0092 & 1.0 & 0.628 & 0.014 & 2.3 & 0.569 & 0.13 & 23 & -1.21 & 0.12 & 10 & 1 & 0.097 & 9.7 \\
 1.4 & 0.911 & 0.010 & 1.1 & 0.613 & 0.020 & 3.3 & 0.654 & 0.11 & 16 & -1.26 & 0.09 & 7.1 & 1 & 0.073 & 7.3 \\
  
\hline \hline
\end{tabular}\caption{Fiducial values and errors for the parameters $P_{1}$, $P_{2}$,
$P_{3}$, $E'/E$ and $\bar\eta$ for every bin. The last bin has been omitted
since $R'$ is not defined there.}
\label{tab:parseta8} 
\end{table}
\endgroup

\begingroup 
\squeezetable
\begin{table}[b]
\centering
\begin{tabular}{cccccccccccccccc}
\hline \hline
$\bar{z}$  & $P_{1}$  & $\Delta P_{1}$  & $\Delta P_{1}(\%)$  &
$P_{2}$  &
$\Delta P_{2}$  & $\Delta P_{2}(\%)$  & $P_{3}$  & $\Delta P_{3}$  & $\Delta
P_{3}(\%)$  & $(E'/E)$  & $\Delta E'/E$ & $\Delta E'/E (\%)$ &
$\bar\eta$  & $\Delta\bar\eta$  & $\Delta\bar\eta(\%)$ \tabularnewline \hline
 0.7 & 0.794 & 0.0079 & 0.99 & 0.703 & 0.0074 & 1.0 & 0.231 & 0.042 & 18 & -0.983 & 0.023 & 2.3 & 1 & 0.031 & 3.1 \\
 1.1 & 0.875 & 0.0067 & 0.77 & 0.638 & 0.0072 & 1.1 & 0.518 & 0.050 & 9.7 & -1.17 & 0.044 & 3.7 & 1 & 0.037 & 3.7 \\
 1.5 & 0.920 & 0.0099 & 1.1 & 0.607 & 0.010 & 1.7 & 0.688 & 0.048 & 7.0 & -1.29 & 0.060 & 4.6 & 1 & 0.032 & 3.2 \\
  
\hline \hline
\end{tabular}\caption{Same as Table~\ref{tab:parseta8}, but with four redshift bins. The
last bin has again been omitted.}
\label{tab:parseta4} 
\end{table}
\endgroup

\section{Combining the matrices\label{sec:comb}}

Once we have the three Fisher matrices for galaxy clustering, weak lensing and supernovae, we insert them block-wise into
a $(4n_{B})\times(4n_{B})$ matrix for the full parameter vector 
\begin{equation}
p_{\alpha}=\{\bar{A},\bar{R},\bar{L},E\}\times n_{B},
\end{equation}
 Notice that we need also $\bar{R}'=-(1+z)[\bar{R}(z+\Delta z)-\bar{R}(z)]/\Delta z$
and $E'=-(1+z)[E(z+\Delta z)-E(z)]/\Delta z$. 
The full schematic
structure \emph{for every bin} will be: 
\begin{eqnarray}
 &  & \left(\begin{array}{cccc}
\bar{A}\bar{A} & \bar{A}\bar{R} & 0 & \bar{A}E\\
\bar{A}\bar{R} & \bar{R}\bar{R} & 0 & \bar{R}E\\
0 & 0 & \bar{L}\bar{L} & \bar{L}E\\
\bar{A}E & \bar{R}E & \bar{L}E & (EE)^{\Sigma}
\end{array}\right),
\end{eqnarray}
with $(EE)^{\Sigma} = (EE)^{\text{GC}}+(EE)^{\text{WL}}+ (EE)^{\text{SN}}$.
This matrix must then be projected onto $\bar\eta$. It is however interesting to
produce two intermediate steps, namely the matrix for
$q_{\alpha}=\{P_{1},P_{2},P_{3},E\}$
where $P_{1}=R/A$, $P_{2}=L/R$ and $P_{3}=R'/R$, as well as the matrix for
$q_\alpha = \{P_1,P_2,P_3,E'/E\}$. They are given by
\begin{equation}
F_{\alpha\beta}^{(q)}=F_{\gamma\delta}^{(p)}\frac{\partial
p_{\gamma}}{\partial q_{\alpha}}\frac{\partial p_{\delta}}{\partial
q_{\beta}}. 
\end{equation}
We then project onto
$\{P_1,P_{2},\bar\eta,E\}$. In Table~\ref{tab:parseta8} we present the fiducial
values for the parameters $P_{1}$, $P_{2}$, $P_{3}$, defined in
Sec.~\ref{sec:int}; In Fig.~\ref{figerrPi} we plot their
fiducial values and errors. Let us call this the basic Fisher matrix.

As we mentioned in the introduction, we decided to consider four
models for $\bar\eta$: constant, variable only in redshift,
variable both in space and redshift, and the Horndeski model. For the constant
$\bar\eta$ case we project the basic Fisher Matrix for
$P_1,P_{2},\bar\eta,E$ onto a single constant value for $\bar\eta$.
The resulting uncertainty for $\bar\eta$ is
0.010
\unskip.

For the $z$-variable case we project on five $\bar\eta$ parameters, one for
each bin. The results are in Table~\ref{tab:parseta8}. We see that the error
on $\bar\eta$ rises to around 10\%.  Without the SN data, the final constraints
on $\eta$ would weaken only by roughly 1\%.  If we collect the data into
only three wider $z$ bins, the error reduces to about 3\%.

For the $z,k$ varying case,  we consider the $k$-binning of
Sec.~\ref{subsec:gck}.  Now the information is distributed over many more
bins, so the errors obviously degrade (see Table~\ref{tab:errorsPetakbins}).
We find errors from 10\% to more than 100\%.

\begin{figure}[t]
\centering
\includegraphics[width=0.6\linewidth]{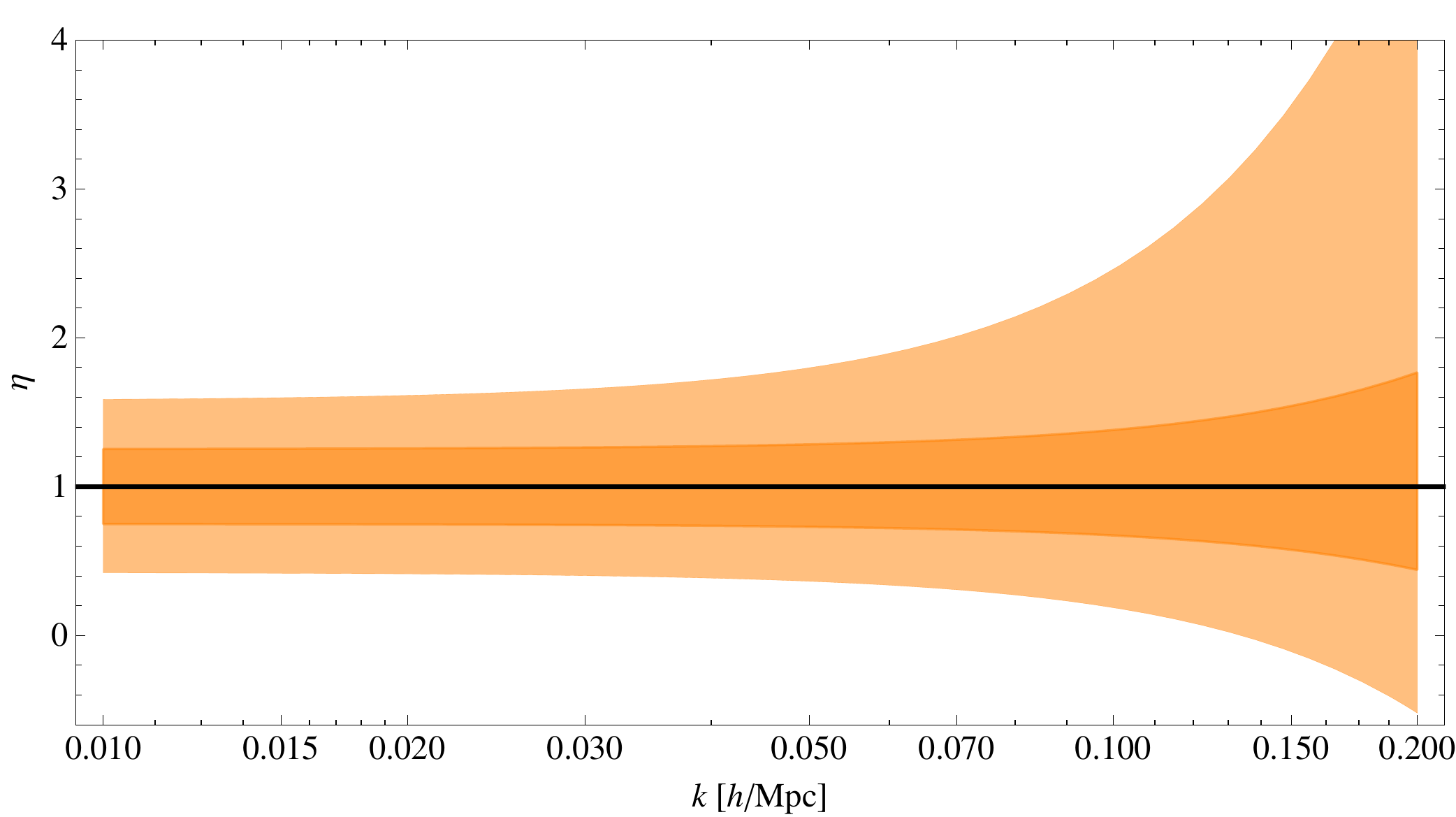} 
\caption{\label{fig:erroretalog} Constraints on $\eta(k)$ in the Horndeski
case for $z=0.6$ (light) and $z=1.4$ (dark).}
\end{figure}

Finally, for the Horndeski case, Table~\ref{tab:h1h3} gives the absolute
errors on $h_2,h_4$ (measuring $k$ in units of 0.1 $h/$Mpc).  Here we are
forced to fix $h_5$ to its fiducial value (i.e. to zero) due to the
degeneracy between $h_4$ and $h_4$ when the fiducial model is  such that
$h_4=h_5$, as in the $\Lambda$CDM case. This means we are only able to measure
the difference $h_4-h_5$ rather than the two functions separately.  The
absolute errors on $h_2,h_4$ are in the range 0.2-0.6. This result implies
for instance that, at a scale of 0.1 $h/$Mpc and in a redshift bin 0.5-0.7,
a Euclid-like mission can detect the presence of a $k^2$ behavior in
$\eta$ if it is larger than 60\% than the $k$-independent trend (see
Fig.~\ref{fig:erroretalog} for a visualization of the constraints on
$\eta$).

\begingroup \squeezetable
\begin{table}[b]
 \begin{tabular}{cccccccccccc}
\hline \hline
\multicolumn{2}{c}{} & \multicolumn{2}{c}{WL} & \multicolumn{2}{c}{GC}
& \multicolumn{2}{c}{SN} & \multicolumn{2}{c}{WL+GC} &
\multicolumn{2}{c}{WL+GC+SN}\tabularnewline \hline 
$\bar{z}$  & $E$  & $\Delta E$  & $\Delta E(\%)$ & $\Delta E$  &
$\Delta E(\%)$ & $\Delta E$  & $\Delta E(\%)$  & $\Delta E$  & $\Delta
E(\%)$ & $\Delta E$  & $\Delta E(\%)$\tabularnewline \hline 
 0.6 & 1.37 & 0.0062 & 0.46 & 0.12 & 8.5 & 0.0026 & 0.19 & 0.0062 & 0.45 & 0.0023 & 0.16 \\
 0.8 & 1.53 & 0.0069 & 0.45 & 0.073 & 4.8 & 0.0041 & 0.27 & 0.0068 & 0.44 & 0.0029 & 0.19 \\
 1.0 & 1.72 & 0.017 & 0.96 & 0.058 & 3.4 & 0.0086 & 0.50 & 0.016 & 0.91 & 0.0067 & 0.39 \\
 1.2 & 1.92 & 0.029 & 1.5 & 0.050 & 2.6 & 0.016 & 0.83 & 0.024 & 1.2 & 0.012 & 0.65 \\
 1.4 & 2.14 & 0.029 & 1.4 & 0.051 & 2.4 & 0.028 & 1.3 & 0.022 & 1.0 & 0.017 & 0.78 \\
 1.8 & 2.62 & 0.077 & 3.0 & 0.061 & 2.3 & - & - & 0.046 & 1.8 & 0.043 & 1.7 \\
  
\hline \hline
\end{tabular}\caption{Errors on $E$ from the three probes.}
\label{tab:allerrorsE} 
\end{table}
\endgroup

\begingroup \squeezetable
\begin{table}[t]
\centering
\begin{tabular}{cccccccccccccc}
\hline \hline
$\bar z$ & $i$ & $P_1$ & $\Delta P_1$ & $\Delta P_1(\%)$ &$P_2$ 
& $\Delta P_2$ & $\Delta P_2(\%)$ &$P_3$ & $\Delta P_3$ & $\Delta P_3(\%)$ 
& $\bar \eta$ & $\Delta \bar \eta$ & $\Delta \bar \eta(\%)$\\
\hline
 \hline\multirow{3}{*}{0.6} & 1 & \multirow{3}{*}{0.766}& 0.14 & 18 & \multirow{3}{*}{0.729}& 0.12 & 17 & \multirow{3}{*}{0.134}& 1.4 & 1100 & \multirow{3}{*}{1}& 1.1 & 120 \\
  & 2 &  & 0.032 & 4.1 &  & 0.030 & 4.1 &  & 0.33 & 240 &  & 0.26 & 26 \\
  & 3 &  & 0.013 & 1.7 &  & 0.015 & 2.0 &  & 0.15 & 110 &  & 0.12 & 12 \\
 \hline\multirow{3}{*}{0.8} & 1 & \multirow{3}{*}{0.819}& 0.11 & 13 & \multirow{3}{*}{0.682}& 0.092 & 13 & \multirow{3}{*}{0.317}& 1.2 & 380 & \multirow{3}{*}{1}& 0.93 & 93 \\
  & 2 &  & 0.024 & 2.9 &  & 0.021 & 3.1 &  & 0.26 & 83 &  & 0.2 & 20 \\
  & 3 &  & 0.011 & 1.4 &  & 0.013 & 1.9 &  & 0.14 & 43 &  & 0.1 & 10 \\
 \hline\multirow{3}{*}{1.0} & 1 & \multirow{3}{*}{0.859}& 0.093 & 11 & \multirow{3}{*}{0.65}& 0.076 & 12 & \multirow{3}{*}{0.46}& 1.1 & 240 & \multirow{3}{*}{1}& 0.82 & 82 \\
  & 2 &  & 0.020 & 2.3 &  & 0.019 & 2.9 &  & 0.23 & 51 &  & 0.17 & 17 \\
  & 3 &  & 0.011 & 1.2 &  & 0.012 & 1.8 &  & 0.14 & 31 &  & 0.1 & 11 \\
 \hline\multirow{3}{*}{1.2} & 1 & \multirow{3}{*}{0.888}& 0.084 & 9.4 & \multirow{3}{*}{0.628}& 0.074 & 12 & \multirow{3}{*}{0.569}& 1.1 & 190 & \multirow{3}{*}{1}& 0.78 & 78 \\
  & 2 &  & 0.017 & 2.0 &  & 0.021 & 3.3 &  & 0.23 & 40 &  & 0.16 & 16 \\
  & 3 &  & 0.011 & 1.2 &  & 0.017 & 2.7 &  & 0.17 & 29 &  & 0.12 & 12 \\
 \hline\multirow{3}{*}{1.4} & 1 & \multirow{3}{*}{0.911}& 0.079 & 8.7 & \multirow{3}{*}{0.613}& 0.084 & 14 & \multirow{3}{*}{0.654}& 0.79 & 120 & \multirow{3}{*}{1}& 0.55 & 55 \\
  & 2 &  & 0.017 & 1.9 &  & 0.027 & 4.4 &  & 0.17 & 26 &  & 0.12 & 12 \\
  & 3 &  & 0.013 & 1.4 &  & 0.023 & 3.8 &  & 0.14 & 21 &  & 0.094 & 9.4 \\

\hline \hline
\end{tabular}
\caption{Here, the errors on $P_1$, $P_2$, $P_3$ and $\eta$ are listed for the $z,k$-varying 
case with a similar structure as Table~\ref{tab:errorsARE}.}
\label{tab:errorsPetakbins}
\end{table}
\endgroup

\begingroup \squeezetable
\begin{table}[b]
\centering
\begin{tabular}{ccc}
\hline \hline
$\bar z$ &  $\Delta h_2$ & $\Delta h_4$  \\ \hline
   0.6 & 0.58 & 0.56 \\
 0.8 & 0.44 & 0.32 \\
 1. & 0.37 & 0.22 \\
 1.2 & 0.35 & 0.18 \\
 1.4 & 0.25 & 0.1 \\
\hline \hline
\end{tabular}
\caption{Absolute errors on $h_2$ and  $h_4$. Because of the
degeneracy between $h_5$ and $h_4$, $h_5$ has been fixed. The fiducial
values are $h_2=1$ and $h_4=0$.}
\label{tab:h1h3}
\end{table}
\endgroup


\section{Conclusions}

In this paper we study the precision with which a future large survey of
galaxy clustering and weak lensing like Euclid can determine the anisotropic
stress of the dark sector with the help of the model-independent
cosmological observables introduced in \cite{Amendola:2012ky}, when
augmented with a supernova survey. 

We find that galaxy clustering and weak lensing will achieve  precise
measurements of the expansion rate $E(z)=H(z)/H_{0}$, with errors of less
than a percent in redshift bins of $\Delta z=0.2$ out to $z=1.5$, and with
less than 4\% out to $z=2$, see Table~\ref{tab:allerrorsE}.

They will also be able to measure $P_{1}=f/b$ to about a percent precision
over the full redshift range (in the same bins), and achieve a comparable
precision on $P_{2}=\Omega_{{\rm m,0}}\Sigma/f$, except at $z>1.5$ where the
errors increase rapidly. The final quantity, $P_{3}=f+f'/f$, is constrained
much less precisely, only to about 30\%,  because it involves an
explicit derivative. The detailed results are given in
Tables~\ref{tab:parseta8} and \ref{tab:parseta4}.

We then considered four different models for $\eta=-\Phi/\Psi$: 
\begin{enumerate}
\item A constant $\eta$: In this case we find that we can determine  the derived quantity $\bar\eta$
with a precision of about 1\%. 
\item $\eta$ varying with redshift, but not with scale: For bins with a
size of $\Delta z=0.2$, we find a precision on $\bar\eta$ of about 10\% out to
$z=1.5$. 
\item $\bar\eta$ varying both in $z$ and in $k$: the errors vary considerably across the $z,k$ range, 
from 10\% to more than 100\%.
\item The Horndeski case: now the absolute errors on $h_2,h_4$ are in the range 0.2-0.6
\end{enumerate}
We stress again that in this paper we used {\em
only} directly observable quantities without any further assumptions
about the initial power spectrum, the dark matter, the dark energy
model (beyond the behaviour of $\eta$ in the last step) or the bias,
as such assumptions may be unwarranted in a general dark energy or
modified gravity context. 
On the other hand, we do assume that a window between 
non-linear scales and sub-sound-horizon scales exists and is wide enough
 to cover all the wavelengths we have been employing in our forecasts.

\begin{acknowledgements} We thank M. Motta, I. Saltas and I. Sawicki for useful discussions. L.A., A.G. and A.V. acknowledge
support from DGF through the project TRR33 ``The Dark Universe''. A.G. also acknowledges financial support from DAAD through program 
``Forschungsstipendium f\"{u}r Doktoranden und Nachwuchswissenschaftler''. M.K. acknowledges financial support from the Swiss NSF.
\end{acknowledgements}

\appendix

\section{Sampling vs Fisher matrix analysis\label{sec:appeta}}

In order to check whether the Fisher matrix analysis is appropriate for the non-linear
parameter combinations that make up the $P_i$ and $\eta$ we also use an alternative
approach. We assume that the Fisher matrix forecast for the errors on $\bar A$, $\bar R$,
$\bar L$ and $E$ is sufficiently accurate (i.e.\ that the joint posterior of these variables can be described by
a Gaussian probability distribution function with the covariance matrix given by the inverse of the Fisher matrix), 
which should be a reasonable assumption
given how precise the surveys that we consider here are. We then draw random samples from the 
multivariate Gaussian distribution defined by those Fisher matrices.

For each sample we compute $P_1$, $P_2$ and $P_3$ at the corresponding values of
$z$ and $k$. We compute the derivatives of $E$ and $\bar R$ by fitting a cubic spline through
each realisation of $E(z)$ and $\bar R(z)$ and calculating the derivative of the spline. This procedure
allows us to obtain estimates of the derivatives in all bins, but at the price of having to choose
boundary conditions for the splines (we use the ``natural spline'' convention that the second derivative
vanishes at the boundary).

\begin{center}
\begin{figure}[t]
\includegraphics[width=0.4\linewidth]{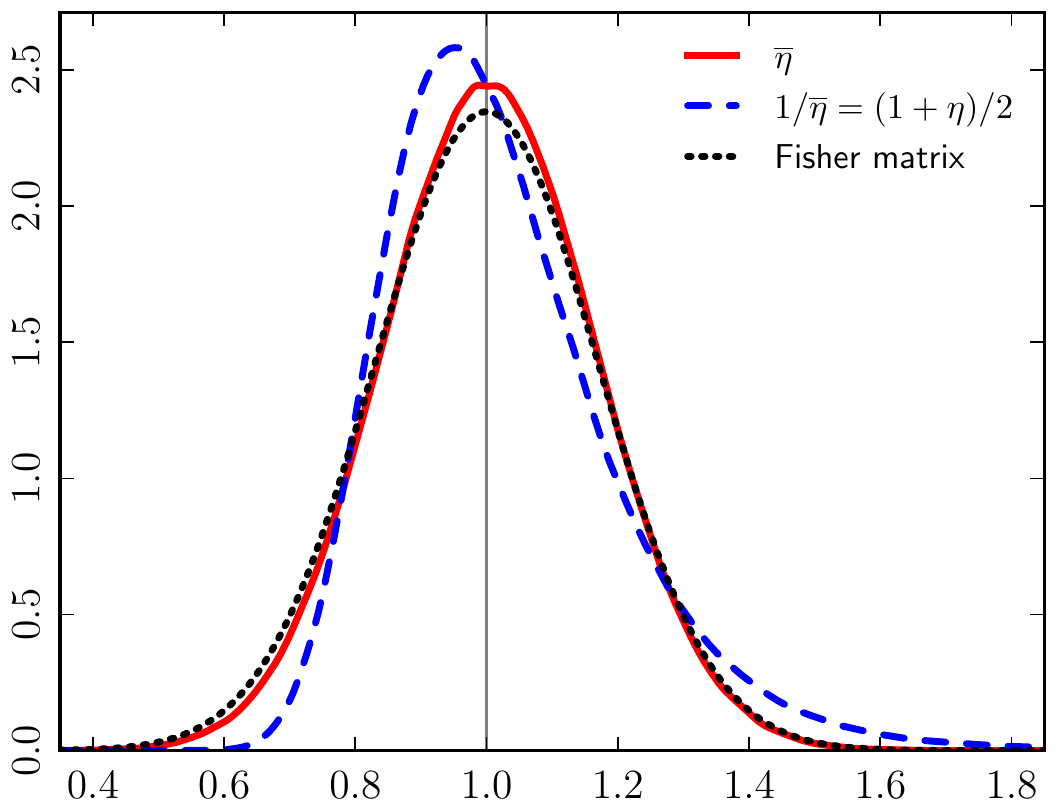}  \quad
\includegraphics[width=0.4\linewidth]{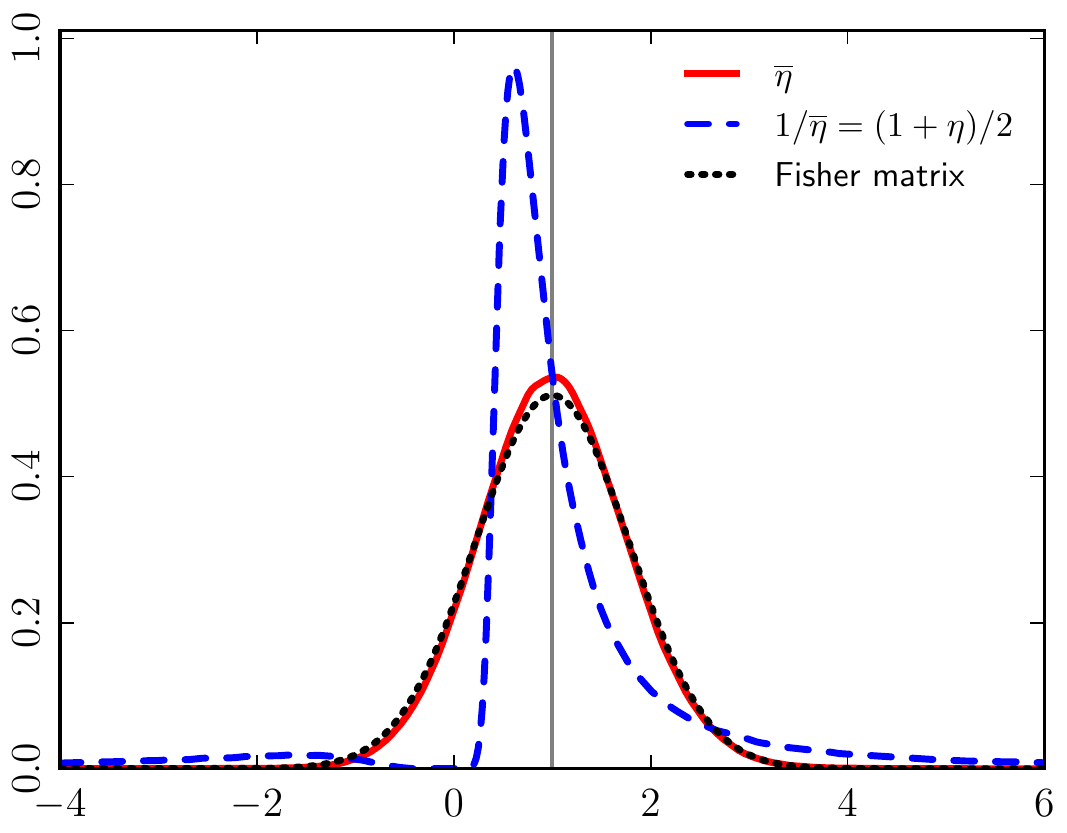}
\caption{\label{fig:sampling_pdf} The probability density function (pdf) for $(1+\eta)/2$ (blue dashed line) and $\bar\eta$ (red solid line) based on sampling
from the Fisher matrix for $\{\bar A, \bar R, \bar L, E\}$, compared to the Gaussian pdf from the Fisher matrix projection on $\bar\eta$
(black dotted line) in the $z$- and $k$-binning case.
The left panel shows the second $k$-bin for $\bar z = 1$, and the right panel the first $k$-bin for $\bar z = 1.2$. We use $(1+\eta)/2$
instead of $\eta$ because it has the same pdf shape as $\eta$ and (to lowest order) the same variance as $\bar\eta$. 
We see that even when the standard deviation of $\bar\eta$ is well below $1$ as in the left panel, the pdf of $\eta$ is significantly
less Gaussian than the pdf of $\bar\eta$. For large standard deviation (right panel) the pdf of $\bar\eta$ is still well behaved and
close to Gaussian, while the one of $\eta$ is strongly distorted and exhibits large tails (not shown in the figure) due to a division 
by zero problem in the expression (\ref{eq:obseta}). }
\end{figure}
\end{center}

Overall we find good agreement, and even excellent agreement when using the derivative at the points
in between the bins (which agrees better with the finite difference method used for the Fisher forecasts).
The agreement becomes much worse for $\eta$, as already mentioned in the introduction. This is
however no surprise, as the posterior distribution of $\eta$ becomes very non-Gaussian for the survey
specifications considered here (while the posterior distributions of the $P_i$ remain close to Gaussian).
We observe however that $\bar\eta$ retains a normal posterior, which makes it much better suited for
the Fisher forecast approach, see Fig.\ \ref{fig:sampling_pdf}. The same holds true for Markov-Chain Monte Carlo approaches which
tend to have difficulties with sampling from curved, ``banana-shaped'' posteriors, and so we recommend
quite generally to use $\bar\eta$ rather than $\eta$ in data analysis. We finally note that when $\eta$ is well-constrained
and has a pdf close to Gaussian, then its standard deviation should be about twice that of $\bar\eta$.

\bibliographystyle{utcaps}
\bibliography{observables}

\providecommand{\href}[2]{#2}\begingroup\raggedright\begin{thebibliography}{10}

\bibitem{Ade:2013ktc}
{\bfseries Planck Collaboration} Collaboration, P.~Ade {\em et~al.}, ``{Planck
  2013 results. I. Overview of products and scientific results},''
\href{http://arxiv.org/abs/1303.5062}{{\ttfamily arXiv:1303.5062
  [astro-ph.CO]}}.

\bibitem{Ade:2013zuv}
{\bfseries Planck Collaboration} Collaboration, P.~Ade {\em et~al.}, ``{Planck
  2013 results. XVI. Cosmological parameters},''
\href{http://arxiv.org/abs/1303.5076}{{\ttfamily arXiv:1303.5076
  [astro-ph.CO]}}.

\bibitem{Amendola:2012ky}
L.~Amendola, M.~Kunz, M.~Motta, I.~D. Saltas, and I.~Sawicki, ``{Observables
  and unobservables in dark energy cosmologies},''
  \href{http://dx.doi.org/10.1103/PhysRevD.87.023501}{{\em Phys.Rev.}
  {\bfseries D87} (2013) 023501},
\href{http://arxiv.org/abs/1210.0439}{{\ttfamily arXiv:1210.0439
  [astro-ph.CO]}}.

\bibitem{Amendola:2007rr}
L.~Amendola, M.~Kunz, and D.~Sapone, ``{Measuring the dark side (with weak
  lensing)},'' \href{http://dx.doi.org/10.1088/1475-7516/2008/04/013}{{\em
  JCAP} {\bfseries 0804} (2008) 013},
\href{http://arxiv.org/abs/0704.2421}{{\ttfamily arXiv:0704.2421 [astro-ph]}}.

\bibitem{DeFelice:2011hq}
A.~De~Felice, T.~Kobayashi, and S.~Tsujikawa, ``{Effective gravitational
  couplings for cosmological perturbations in the most general scalar-tensor
  theories with second-order field equations},''
  \href{http://dx.doi.org/10.1016/j.physletb.2011.11.028}{{\em Phys.Lett.}
  {\bfseries B706} (2011) 123--133},
\href{http://arxiv.org/abs/1108.4242}{{\ttfamily arXiv:1108.4242 [gr-qc]}}.

\bibitem{Saltas:2010tt}
I.~D. Saltas and M.~Kunz, ``{Anisotropic stress and stability in modified
  gravity models},'' \href{http://dx.doi.org/10.1103/PhysRevD.83.064042}{{\em
  Phys.Rev.} {\bfseries D83} (2011) 064042},
\href{http://arxiv.org/abs/1012.3171}{{\ttfamily arXiv:1012.3171 [gr-qc]}}.

\bibitem{Sawicki:2012re}
I.~Sawicki, I.~D. Saltas, L.~Amendola, and M.~Kunz, ``{Consistent perturbations
  in an imperfect fluid},''
  \href{http://dx.doi.org/10.1088/1475-7516/2013/01/004}{{\em JCAP} {\bfseries
  1301} (2013) 004},
\href{http://arxiv.org/abs/1208.4855}{{\ttfamily arXiv:1208.4855
  [astro-ph.CO]}}.

\bibitem{Motta:2013cwa}
M.~Motta, I.~Sawicki, I.~D. Saltas, L.~Amendola, and M.~Kunz, ``{Probing Dark
  Energy through Scale Dependence},''
\href{http://arxiv.org/abs/1305.0008}{{\ttfamily arXiv:1305.0008
  [astro-ph.CO]}}.

\bibitem{Kunz:2007rk}
M.~Kunz, ``{The dark degeneracy: On the number and nature of dark
  components},'' \href{http://dx.doi.org/10.1103/PhysRevD.80.123001}{{\em
  Phys.Rev.} {\bfseries D80} (2009) 123001},
\href{http://arxiv.org/abs/astro-ph/0702615}{{\ttfamily arXiv:astro-ph/0702615
  [astro-ph]}}.

\bibitem{Ballesteros:2011cm}
G.~Ballesteros, L.~Hollenstein, R.~K. Jain, and M.~Kunz, ``{Nonlinear
  cosmological consistency relations and effective matter stresses},''
  \href{http://dx.doi.org/10.1088/1475-7516/2012/05/038}{{\em JCAP} {\bfseries
  1205} (2012) 038},
\href{http://arxiv.org/abs/1112.4837}{{\ttfamily arXiv:1112.4837
  [astro-ph.CO]}}.

\bibitem{Albrecht:2006um}
A.~Albrecht, G.~Bernstein, R.~Cahn, W.~L. Freedman, J.~Hewitt, {\em et~al.},
  ``{Report of the Dark Energy Task Force},''
\href{http://arxiv.org/abs/astro-ph/0609591}{{\ttfamily arXiv:astro-ph/0609591
  [astro-ph]}}.

\bibitem{2011arXiv1110.3193L}
R.~{Laureijs}, J.~{Amiaux}, S.~{Arduini}, J.~. {Augu{\`e}res}, J.~{Brinchmann},
  R.~{Cole}, M.~{Cropper}, C.~{Dabin}, L.~{Duvet}, A.~{Ealet}, and et~al.,
  ``{Euclid Definition Study Report},'' {\em ArXiv e-prints} (Oct., 2011) ,
  \href{http://arxiv.org/abs/1110.3193}{{\ttfamily arXiv:1110.3193
  [astro-ph.CO]}}.

\bibitem{Tyson:2003kb}
{\bfseries LSST Collaboration} Collaboration, J.~A. Tyson, ``{Large synoptic
  survey telescope: Overview},'' {\em Proc.SPIE Int.Soc.Opt.Eng.} {\bfseries
  4836} (2002) 10--20,
\href{http://arxiv.org/abs/astro-ph/0302102}{{\ttfamily arXiv:astro-ph/0302102
  [astro-ph]}}.

\bibitem{Seo:2003}
H.-J.~J. Seo and D.~J. Eisenstein, ``{Probing dark energy with baryonic
  acoustic oscillations from future large galaxy redshift surveys},''
  \href{http://dx.doi.org/10.1086/379122}{{\em Astrophys.J} {\bfseries 598}
  (2003) 720--740},
\href{http://arxiv.org/abs/0307460}{{\ttfamily arXiv:0307460 [astro-ph]}}.

\bibitem{Alcock:1979}
C.~{Alcock} and B.~{Paczynski}, ``{An evolution free test for non-zero
  cosmological constant},'' \href{http://dx.doi.org/10.1038/281358a0}{{\em
  \nat} {\bfseries 281} (Oct., 1979) 358}.

\bibitem{Amendola2010}
L.~Amendola and S.~Tsujikawa, {\em {Dark Energy: Theory and Observations}}.
\newblock Cambridge University Press, 2010.

\bibitem{Amendola:2012ys}
{\bfseries Euclid Theory Working Group} Collaboration, L.~Amendola {\em
  et~al.}, ``{Cosmology and fundamental physics with the Euclid satellite},''
\href{http://arxiv.org/abs/1206.1225}{{\ttfamily arXiv:1206.1225
  [astro-ph.CO]}}.

\bibitem{Lewis:2000}
A.~Lewis, A.~Challinor, and A.~A. Lasenby, ``{Efficient Computation of CMB
  anisotropies in closed FRW models},''
  \href{http://dx.doi.org/10.1086/309179}{{\em Astrophys.J.} {\bfseries 538}
  (2000) 473--476}, \href{http://arxiv.org/abs/9911177}{{\ttfamily
  arXiv:9911177 [astro-ph]}}.
Code available in \url{http://camb.info/}.

\bibitem{Orsi:2009mj}
A.~Orsi, C.~Baugh, C.~Lacey, A.~Cimatti, Y.~Wang, {\em et~al.}, ``{Probing dark
  energy with future redshift surveys: A comparison of emission line and broad
  band selection in the near infrared},''
\href{http://arxiv.org/abs/0911.0669}{{\ttfamily arXiv:0911.0669
  [astro-ph.CO]}}.

\bibitem{Hu:1999ek}
W.~Hu, ``{Power spectrum tomography with weak lensing},''
  \href{http://dx.doi.org/10.1086/312210}{{\em Astrophys.J.} {\bfseries 522}
  (1999) L21--L24},
\href{http://arxiv.org/abs/astro-ph/9904153}{{\ttfamily arXiv:astro-ph/9904153
  [astro-ph]}}.

\bibitem{Ma2006}
Z.~{Ma}, W.~{Hu}, and D.~{Huterer}, ``{Effects of Photometric Redshift
  Uncertainties on Weak-Lensing Tomography},''
  \href{http://dx.doi.org/10.1086/497068}{{\em \apj} {\bfseries 636} (Jan.,
  2006) 21--29}, \href{http://arxiv.org/abs/0506614}{{\ttfamily arXiv:0506614
  [astro-ph]}}.

\bibitem{Eisenstein1999a}
D.~J. {Eisenstein}, W.~{Hu}, and M.~{Tegmark}, ``{Cosmic Complementarity: Joint
  Parameter Estimation from Cosmic Microwave Background Experiments and
  Redshift Surveys},'' \href{http://dx.doi.org/10.1086/307261}{{\em \apj}
  {\bfseries 518} (June, 1999) 2--23},
  \href{http://arxiv.org/abs/astro-ph/9807130}{{\ttfamily astro-ph/9807130}}.

\end{thebibliography}\endgroup

\end{document}